\documentclass[12pt]{article}
\usepackage{graphicx}

\usepackage{epsfig}
\usepackage{amsmath}
\usepackage{amssymb}
\usepackage{amsthm}
\usepackage{latexsym}

\newcommand{\comment}[1]{}
\newtheorem{theorem}{Theorem}
\newtheorem{lemma}{Lemma}[section]
\newtheorem{remark}{Remark}[section]
\newtheorem{corollary}{Corollary}[section]

\newtheorem{definition}{Definition}[section]
\begin{document}

\title{{\LARGE\sf
{\bf Universality in Two-Dimensional Enhancement Percolation}}
}

\author{
{\bf Federico Camia}
\thanks{Department of Mathematics, Vrije Universiteit Amsterdam,
De Boelelaan 1081a, 1081 HV Amsterdam, The Netherlands.
E-mail: fede\,@\,few.vu.nl}
\thanks{Partially supported by a VENI grant of the NWO.}\\
}

\date{}

\maketitle

\begin{abstract}
%
%
We consider a type of dependent percolation introduced
in~\cite{ag}, where it is shown that certain ``enhancements" of
independent (Bernoulli) percolation, called essential, make the
percolation critical probability strictly smaller.
In this paper we first prove that, for two-dimensional enhancements
with a natural monotonicity property, being essential is also a
necessary condition to shift the critical point.
We then show that (some) critical exponents and the scaling limit of
crossing probabilities of a two-dimensional percolation process are
unchanged if the process is subjected to a monotonic enhancement
that is not essential. This proves a form of universality for all
dependent percolation models obtained via a monotonic enhancement
(of Bernoulli percolation) that does not shift the critical point.
For the case of site percolation on the triangular lattice,
we also prove a stronger form of universality by showing that
the \emph{full} scaling limit~\cite{cn2,cn3} is not affected by
any monotonic enhancement that does not shift the critical point.
\end{abstract}



\section{Introduction}

One of the most interesting phenomena in statistical physics
is the presence, in the phase diagram of many physical systems,
of ``critical points'' where some thermodynamic quantities or
their derivatives diverge.
Experimentally, it is found that such divergences are usually
of power law type, and therefore characterized by exponents,
called critical exponents.
The theory of critical phenomena, developed to explain this
behavior, suggests the existence of large ``universality classes''
such that systems belonging to the same universality class
have the same critical exponents.

A closely related notion of universality concerns the
``continuum scaling limit,'' which is obtained by sending the
\emph{microscopic scale} of the system (e.g., the mesh of the
lattice, for discrete systems defined on a lattice) to zero,
while focusing on features manifested on a \emph{macroscopic scale}.
Such a limit is only meaningful at the critical point, where the
\emph{correlation length} (i.e., the ``natural" length scale that
characterizes the system) is supposed to diverge.

The concept of universality and the existence of universality
classes arise naturally in the theory of critical phenomena (based
on Renormalization Group techniques), and are backed by strong
theoretical and experimental evidence.
Nonetheless, very few rigorous results are available, especially
below the upper critical dimension, where the values of the critical
exponents are expected to be different from those predicted by mean-field
theory (there are, however, some exceptions -- see, e.g.,
\cite{wierman1,wierman2,spencer,cns1,cns2,cn1,camia,ps,cl,bcl}
for results concerning percolation and the Ising model).
The main aim of this paper is to present some rigorous results in
support of the idea of universality in the context of percolation
theory in two dimensions (see~\cite{kesten,grimmett} for detailed
accounts on percolation theory).

In recent years, substantial progress has been made in understanding
two-dimensional critical percolation and its conformally invariant
scaling limit (see, e.g., \cite{smirnov,smirnov-long,schramm1,dubedat1,cn2,dubedat2,cn3,cn4},
and in computing the percolation critical exponents by means of
mathematically rigorous methods~\cite{sw,lsw}.
The new tool that made this possible is the Stochastic Loewner
Evolution introduced by Schramm~\cite{schramm}, together with
Smirnov's proof~\cite{smirnov,smirnov-long} of the conformal
invariance of the scaling limit of crossing probabilities for
site percolation on the triangular lattice.

Here we will focus on a class of percolation models, called enhancement
percolation, introduced in~\cite{ag} (see also~\cite{grimmett}).
Enhancement percolation configurations are obtained by modifying,
according to some local set of rules, the configurations generated
by an independent (Bernoulli) percolation process.
Other dependent percolation models obtained in a similar way arise
naturally in different contexts such as the nonequilibrium dynamics
of stochastic Ising models at zero temperature (see, for instance,
\cite{bray,gns,howard,nns,ns,cdn,fss}) and cellular automata like
bootstrap percolation (see, for instance, \cite{al,schonmann,fis}).


The enhancements that we are interested in are endowed with a natural
monotonicity property and are such that they do not change the nature
of the phase transition and do not shift the critical point $p_c$, therefore
transforming an independent critical percolation model into a different,
but still critical, dependent percolation model.
However, even an enhancement that does not shift $p_c$ can modify
significantly the initial percolation process, and it could a priori
induce ``macroscopic'' changes; it is therefore natural to ask whether
such an enhancement can change the continuum scaling limit and the
critical exponents.
We show that this is not the case for the class of enhancements studied
here.

The result is not surprising, if seen in the context of the Renormalization
Group picture which suggests that, as long as the enhancement has
finite range, it should not affect the scaling limit and the critical
exponents (see, for example, \cite{cardy2}).
However, the Renormalization Group picture remains largely nonrigorous and
no general formalism has been developed so far to put it on more solid ground.
Moreover, as the eight-vertex model~\cite{baxter} and some spin
models~\cite{leeuwen,mastropietro1,mastropietro2} show, the concept
of universality needs to be taken with some care.


Universality results in the same spirit as those presented in this
paper can be found in~\cite{cns1,cn1,cns2,camia} (in~\cite{hn},
universality for critical exponents is tested numerically).
Generally speaking, one way to look at these results is as an attempt
to test the ``robustness'' of the scaling limit and critical exponents,
as well as to give examples of a strong form of universality by constructing
different (dependent) percolation models with the same critical exponents
and scaling limit as independent percolation.
The percolation models considered in~\cite{cns1,cn1,cns2} (and in~\cite{hn})
are generated by applying certain cellular automata to independent
percolation configurations.
In those cases, the initial percolation configurations and the rules
of the cellular automata possess a global ``spin-flip" symmetry.
The situation in~\cite{camia} and in this paper is different because
the dynamics (or the enhancement) breaks the global ``spin-flip" symmetry.

A key tool in proving the universality results is a coupling
between the independent percolation process (before the enhancement)
and the enhanced one, which allows to compare the two processes.
This method does not apply directly to essential enhancements that
do shift the critical point (because in that case, if one starts
at the critical point, the enhanced process is supercritical).
That situation is very interesting and deserves to be studied,
but unfortunately the methods used in this paper do not seem
to be useful there.

Before addressing the universality issue, we present a new result
about two-dimensional enhancements endowed with a natural monotonicity
property.
This complements one of the main results of~\cite{ag}, where it is
proved that certain enhancements, called essential, always shift
the critical point $p_c$.
We show that a two-dimensional, monotonic enhancement that is not
essential cannot produce an infinite cluster, and therefore leaves
the critical point unchanged.
In particular, this means that the monotonic, nonessential
enhancement of a critical (Bernoulli) percolation process is
still critical, and that the phase transition in the enhanced model
is still second order (or continuous).
This motivates the rest of the paper, since it identifies a class
of critical, dependent percolation models for which it is natural
to ask about critical exponents and scaling limits.

The rest of the paper is organized as follows.
The next section contains a preview of some of the main universality
results, in the context of site percolation on the triangular lattice.
Analogous universality results for the square and the hexagonal
lattice are stated later on, in Section~\ref{universality}. Before
that, enhancement percolation is precisely defined and discussed in
Section~\ref{enhancement}, where new results about monotonic
enhancements in two dimensions are also presented. The last two
sections are dedicated respectively to the proofs of the results on
enhancement percolation (Section~\ref{proof-equivalence}) and of the
universality results (Section~\ref{proofs}).


\section{Preview of the Universality Results} \label{preview}

In this section, we present some of the main results of the paper.
For a precise definition of enhancement percolation and more details,
we refer to Section~\ref{enhancement} below.
The universality results collected in this section are limited for
simplicity to the case of the triangular lattice, where they can be
stated unconditionally since the existence of (certain) critical exponents,
of the scaling limit of crossing probabilities, and of the full scaling
limit have been rigorously proved.
Later, in Section~\ref{universality}, we will state some results for
the square and the hexagonal lattice, which will be however conditional
on the existence of the critical exponents and the scaling limit for
independent (site) percolation on those lattices.

\subsection{Monotonic Enhancements} \label{mono-enh}


Consider a dependent (site) percolation model on a regular lattice
$\mathbb L$ in which an initial configuration is generated by
independent variables having density $p$ and then enhanced by means
of a local function of the configuration. By regular lattices we
mean the class of infinite graphs considered in~\cite{kesten}, but
except for Appendix~\ref{matching}, for simplicity we will restrict
our attention to the square, triangular and hexagonal lattice. The
enhancement is stochastically activated at each site with
probability $s$, independently of the other sites, and its effect is
to (possibly) make certain closed sites open. Let $\theta(p,s)$ be
the percolation probability of the enhanced process, i.e., the
probability that the origin belongs to an infinite open cluster
after the enhancement.

In this paper, we restrict our attention to a class of enhancements
endowed with a natural \emph{monotonicity} property, i.e., we
consider enhancement functions that are nondecreasing in the number
of open sites (see Section~\ref{def+main-res} for a precise
definition).
%

While we postpone the precise definitions to
Section~\ref{def+main-res}, we give here a simple example. On a
lattice $\mathbb L$ of degree $D$, consider an enhancement that,
when activated at the origin, makes it open if at least $m \leq D$
of its neighbors are open. This can be considered the prototypical
example of the type of finite-range, monotonic enhancements that we
are interested in. All enhancements in this paper have finite-range
in the sense that, as in the example above, the effect of the
enhancement depends only on a bounded subset of the percolation
configuration around the location where the enhancement is
activated. The enhancement in the example is monotonic in the sense
that making more sites open in the original percolation
configuration can only produce more open sites, and never inhibits
the enhancement.

We remark that the restriction to monotonic enhancements is very natural,
since if the enhancement is not monotonic, in general the percolation
probability $\theta(p,s)$ will not be monotonic in $p$, and there could be
ambiguity over the correct definition of the critical point (i.e., there
could be more than one critical point---as an example, consider an enhancement
of site percolation on the square lattice such that a closed site is made
open only if all of its neighbors are closed).


The following question motivates the rest of the paper. What happens
to the phase transition when a monotonic enhancement is applied at
$p_c$? With regard to this question, there could a priori be three
types of monotonic enhancements: (1) enhancements that make the
percolation process supercritical and therefore shift the critical
point, (2) enhancements that do not make the process supercritical
but change the universality class of the percolation model, (3)
enhancements that do not make the process supercritical and do not
change the universality class of the model. In what follows, we will
essentially show that in two dimensions the second class of
enhancements is empty. In other words, all two-dimensional dependent
percolation models generated by a monotonic enhancement acting on
independent (Bernoulli) percolation and which does not shift the
critical point are in the same universality class as independent
percolation.

In dimension two, we also show (see Theorem~\ref{necess_and_suffic} in
Section~\ref{def+main-res}) that the only monotonic enhancements that
can shift the critical point are those that Aizenman and Grimmett~\cite{ag}
call \emph{essential} (see Section~\ref{def+main-res} for the definition).

\subsection{Critical Exponents} \label{critexp}

One of our universality results concerns four critical exponents, namely the
exponents $\beta$ (related to the percolation probability), $\nu$ (related
to the correlation length), $\eta$ (related to the connectivity function)
and $\gamma$ (related to the mean cluster size).
The existence of these exponents has been recently proved (see~\cite{sw,lsw}
for the details), and their predicted values confirmed rigorously, for the
case of independent site percolation on the triangular lattice $\mathbb T$.
Such exponents are believed to be universal for independent percolation on
regular lattices in the sense that their value should depend only on the
number of dimensions and not on the structure of the lattice or on the nature
of the percolation model (e.g., whether it is site or bond percolation); that
type of universality has not yet been proved.

Consider an independent percolation model on a regular lattice $\mathbb L$
with configurations chosen from a Bernoulli product measure $P_p$ with density
of open sites $p$ ($E_p$ will denote expectation with respect to $P_p$).
Assume that $\mathbb L$ is such that $0<p_c<1$.
Let $C$ be the open cluster containing the origin and $||C||$ its cardinality,
then $\theta(p) = P_p(||C|| = \infty)$ is the {\bf percolation probability}.
Arguments from theoretical physics suggest that $\theta(p)$ behaves roughly like
$(p-p_c)^{\beta}$ as $p$ approaches $p_c$ from above.

It is also believed that the {\bf connectivity function}
\begin{equation} \label{connect-function}
\tau_p(x) = P_p(\text{the origin and } x \text{ belong to the same open cluster})
\end{equation}
behaves, for large Euclidean norm $|x|$, like $|x|^{-\eta}$ if $p=p_c$, and like
$\exp{(-|x|/\xi(p))}$ if $0<p<p_c$, for some $\xi(p)$ satisfying $\xi(p) \to \infty$
as $p \uparrow p_c$.
The {\bf correlation length} $\xi(p)$ is defined by
\begin{equation} \label{correlation-length}
\xi(p)^{-1} = \lim_{|x| \to \infty} \left\{ - \frac{1}{|x|} \log \tau_p(x) \right\}.
\end{equation}
$\xi(p)$ is believed to behave like $(p_c-p)^{-\nu}$ as $p \uparrow p_c$.
The {\bf mean cluster size} $\chi(p) = E_p ||C||$ is also believed to diverge
with a power law behavior $(p_c - p)^{- \gamma}$ as $p \uparrow p_c$.

One possible way to state these conjectures is the following:
\begin{eqnarray}
\lim_{p \downarrow p_c} \frac{\log \theta(p)}{\log (p-p_c)} = \beta, \\
\lim_{|x| \to \infty} \frac{\log \tau_{p_c}(x)}{\log |x|} = - \eta, \\
\lim_{p \uparrow p_c} \frac{\log \xi(p)}{\log (p_c - p)} = - \nu, \\
\lim_{p \uparrow p_c} \frac{\log \chi(p)}{\log (p-p_c)} = - \gamma.
\end{eqnarray}

We call $\theta(p,s)$, $\xi(p,s)$, $\tau_{p,s}(x)$ and $\chi(p,s)$
the quantities analogous respectively to $\theta(p)$, $\xi(p)$,
$\tau_p(x)$ and $\chi(p)$ for the enhanced process (with density
of enhancement $s$).
In the case of site percolation on the triangular lattice, we have
the following result.

\begin{theorem} \label{critexp-triangular}
For every monotonic enhancement (of independent site percolation) on the
triangular lattice that does not shift the critical point, the critical
exponents $\beta$, $\eta$, $\nu$ and $\delta$ exist for the enhanced
percolation process and have the same numerical values as for the
independent process (before the enhancement).
\end{theorem}

In Section~\ref{univ-critexp} below, we will state a similar result for the
square and the hexagonal lattice (Theorem~\ref{cor-critexp}), conditional
on the existence of the critical exponents for independent (site) percolation
on those lattices.


\subsection{Cardy's Formula and the Full Scaling Limit} \label{cardy+full}

Consider the rescaled triangular lattice $\delta{\mathbb T}$.
As the lattice spacing $\delta$ goes to zero, the limit of the probability
of an open crossing in an arbitrary domain between two (distinct) selected
portions of its boundary has been shown~\cite{smirnov} to exist and to be
a conformal invariant of the domain and the two portions of boundary.
This allows to obtain a formula~\cite{cardy1} for crossing probabilities,
first derived by Cardy using nonrigorous methods and bearing his name.

We will show that a monotonic nonessential enhancement does not change
the scaling limit of crossing probabilities.
In particular, this implies the following result.

\begin{theorem} \label{cardy}
For every monotonic enhancement (of independent site percolation) on the
triangular lattice that does not shift the critical point, the crossing
probabilities of the enhanced process converge in the scaling limit to
Cardy's formula.
\end{theorem}

An analogous results for site percolation on the square and hexagonal
lattices, but conditional on the existence of the scaling limit of
crossing probabilities there, is given in Section~\ref{crossing-prob}
(see Theorem~\ref{cross-prob}).

Crossing probabilities only give partial information on a percolation
model and on its scaling limit $\delta \to 0$.
One way to go beyond crossing probabilities is by considering the law
of the random interfaces, along the edges of the dual lattice
$\delta{\mathbb H}$, between open and closed clusters, as suggested by
Schramm~\cite{schramm} (see Figure~\ref{percolation} in Section~\ref{universality-scaling}).
The existence of the scaling limit of the collection of all interfaces
and some of its properties have been derived in~\cite{cn3}, where the
limiting object is called the percolation \emph{full} scaling limit.

We will show that a monotonic nonessential enhancement does not change
the full scaling limit. In particular, this implies the following result.

\begin{theorem} \label{full-triangular}
For every monotonic enhancement (of independent site percolation) on
the triangular lattice that does not shift the critical point, the full
scaling limit of the enhanced process is the same as the full scaling
limit of the independent process.
\end{theorem}

\section{Enhancement Percolation} \label{enhancement}

Consider a dependent percolation process in which the initial
configuration, generated by independent variables having density
$p$, is enhanced by means of a local function of the configuration.
In \cite{ag}, Aizenman and Grimmett ask under which circumstances
the new critical density differs in value from the critical
density $p_c$ of the initial independent percolation.
Looking at site percolation on the $d$-dimensional cubic lattice
as a prototypical example, they introduce a general approach to answer
that question, and give a sufficient condition for the enhancement
to be capable of shifting the critical point.
An analogous question is relevant for all models, such as Ising ferromagnets
and the contact process, that are endowed with certain monotonicity
properties with respect to the critical point (e.g., the addition
of ferromagnetic couplings can only increase the transition temperature).

Loosely speaking, an enhancement is a systematic addition of
open sites performed by means of a translation-invariant
procedure with local rules; if it is capable of creating a
percolation ``backbone," i.e., a doubly-infinite open path,
then it is called \emph{essential}.
A main result of \cite{ag} is that, if $0 < p_c < 1$, an
essential enhancement always shifts the critical point.

Clearly, not all enhancements are essential.
As an example of a nonessential enhancement consider
the addition of an open site at $x$ with probability
$\frac{1}{2} p_c$ whenever all the neighbors of $x$ are closed.
Such an enhancement introduces new open sites, but it cannot
produce a doubly-infinite open path (almost surely).


\subsection{The Lattices and Some Notation} \label{lattices}

We set up here the notation needed in the following sections.
We will state most of our results for three planar lattices, the
square, triangular and hexagonal lattice. The triangular and
hexagonal lattices will be denoted by $\mathbb T$ and $\mathbb H$,
respectively. In Section~\ref{universality-scaling}, however, we
will restrict our attention to site percolation on the triangular
lattice only. Rather than treating the three lattices separately, we
provide a unified treatment which in fact allows for even greater
generality.

\begin{remark}
We note that the results of Sections~\ref{def+main-res}, \ref{trick}
and~\ref{univ-critexp} apply to general regular lattices of the type
considered by Kesten in~\cite{kesten} (see Chapter~2 of~\cite{kesten}
and Appendix~\ref{matching}).
\end{remark}

\begin{figure}[!ht]
\begin{center}
\includegraphics[width=7cm]{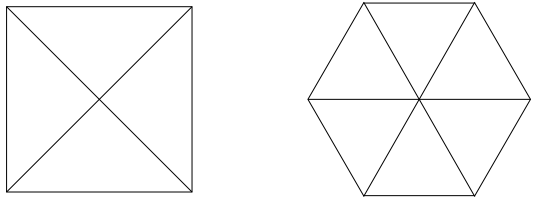}
\caption{The close-packing of the elementary cells of the square
and of the hexagonal lattice.}
\label{close-pack}
\end{center}
\end{figure}
\begin{figure}[!ht]
\begin{center}
\includegraphics[width=7cm]{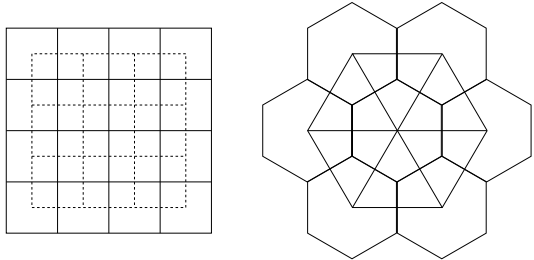}
\caption{As shown on the left, the square lattice is self-dual.
The duality between the triangular and the hexagonal lattice
is shown on the right.}
\label{duality}
\end{center}
\end{figure}

Let $\mathbb L$ be either the square, triangular or hexagonal lattice,
embedded in ${\mathbb R}^2$ as in Figures~\ref{close-pack} and~\ref{duality}.
We think of a lattice as a geometric object made of sites and edges,
and denote by $V({\mathbb L})$ the set sites of $\mathbb L$.
If $F$ a face of $\mathbb L$, we call the {\bf perimeter} of $F$ the
set of edges delimiting $F$, and denote by $V(F)$ the set of sites
of $\mathbb L$ along the perimeter of $F$.

{\bf Close-packing} a face $F$ of $\mathbb L$ means adding an
edge between each pair of vertices of $F$ that do not already
share an edge. In close-packing a face $F$, we shall choose to
draw the new edges inside $F$, as in Figure~\ref{close-pack}.
The lattice ${\mathbb L}^*$ (the {\bf matching graph} or
{\bf close-packed} version of $\mathbb L$) is obtained from
$\mathbb L$ by close-packing all its faces. Note that in the
case of the triangular lattice (or any ``triangulated" lattice),
the close-packed version of the lattice coincides with the
original lattice. Such a lattice is called {\bf self-matching}.

We also introduce the {\bf dual lattice} ${\mathbb L}_d$, whose
sites (called {\bf dual sites}) are the (centers of the) faces of
$\mathbb L$. Two dual sites are neighbors when the perimeters of the
corresponding faces have a common edge. We embed ${\mathbb L}_d$ in
${\mathbb R}^2$ in such a way that each {\bf dual edge} crosses an
edge of $\mathbb L$ (see Figure~\ref{duality}), and denote by
$\text{e}^d_{x,y} = (x,y)_d$ the edge dual to the edge
$\text{e}_{x,y} = (x,y)$ of ${\mathbb L}$ (note that the sites that
appear in this notation are not the dual sites on which the edge is
incident). It is easy to see (Figure~\ref{duality}) that there is a
duality relation between the triangular and the hexagonal lattice.
We will use it in dealing with the full scaling limit of critical
site percolation on $\mathbb T$.

Two sites that are neighbors in $\mathbb L$ (respectively,
${\mathbb L}^*$ or ${\mathbb L}_d$) will also be called
$\mathbb L$-{\bf adjacent} (resp., ${\mathbb L}^*$- or
${\mathbb L}_d$-{\bf adjacent}).
Similarly, two subsets of $\mathbb L$ are said to be
$\mathbb L$-{\bf adjacent} (resp., ${\mathbb L}^*$- or
${\mathbb L}_d$-{\bf adjacent}) if the first one contains
at least one site that is $\mathbb L$-adjacent (resp.,
${\mathbb L}^*$- or ${\mathbb L}_d$-adjacent) to a site
of the second one.

An $\mathbb L$-{\bf path} (resp., $*$-{\bf path} or {\bf dual path})
will be an alternating sequence of $\mathbb L$-adjacent (resp.,
${\mathbb L}^*$- or ${\mathbb L}_d$-adjacent) sites and the
edges between them. The set of sites of a path $\gamma$ will
be denoted by $V(\gamma)$.
If the path is closed, in the sense that the initial and
final sites coincide, it will be called a {\bf loop}.
Sometimes we will simply use the term path, without any
specification, if there is no risk of confusion.
A set $C \in V({\mathbb L})$ is $\mathbb L$-{\bf connected}
(resp.,$*$-{\bf connected}) if $\forall x,y \in C$,
there exists an $\mathbb L$-path (resp., $*$-path)
from $x$ to $y$ that uses only sites in $C$.

Sites of $\mathbb L$ and ${\mathbb L}^*$ will be denoted
by the Latin letters $x$, $y$ and $z$, with the origin
denoted by $o$, and paths on the two lattices by $\gamma$
or $\lambda$ and ${\gamma}^*$ or $\lambda^*$, respectively.
Dual sites will be denoted by the Greek letter $\xi$.

A self-avoiding loop $J$ (i.e., a loop that does not have
self-intersections) is a Jordan curve.
Therefore, by the Jordan theorem, ${\mathbb R}^2 \setminus J$
consists of a bounded component, denoted by $\text{int}(J)$,
and an unbounded component, denoted by $\text{ext}(J)$.
Notice that, in the case of a planar lattice, any site
self-avoiding path is self-avoiding.
A standard loop-removal procedure to extract from a
generic path a site self-avoiding subpath with the same
initial and final sites is described in Chapter~2~of~\cite{kesten}.

The {\bf external (site) boundary} of a set $C \in V({\mathbb L})$
is the set of sites of $\mathbb L$ that are not in $C$ but have
at least one neighbor in $C$.
It is a key observation (see, for example, \cite{se} and
Corollary~2.2~of~\cite{kesten}) that the external (site)
boundary of a nonempty, bounded, $\mathbb L$-connected set
$C$ of sites of $\mathbb L$ forms, together with the edges
between sites in the boundary, a self-avoiding $*$-loop $\lambda^*$
such that all the sites in $C$ belong to $\text{int}(\lambda^*)$.
Note also that any nonempty, bounded, $\mathbb L$-connected set
of sites of $\mathbb L$ is surrounded by a self-avoiding dual loop.

\subsection{Essential and Monotonic Enhancements} \label{def+main-res}

Although the choice of bond or site percolation is irrelevant
for our purposes, we will consider, for definiteness, a site
percolation model on ${\mathbb L}$ with configuration
$\eta \in \{ 0, 1 \}^{V({\mathbb L})}$ chosen from a Bernoulli
product measure $P_p$ with density of open sites $p$ ($E_p$ will
denote expectation with respect to $P_p$).
One reason for dealing with site percolation is that every bond
model can be reformulated as a site model on a different lattice
(the converse is not true and therefore site models are more general
than bond models -- for more details, see~\cite{kesten}).
We interpret the value $\eta(x) = 1$ as meaning that $x \in V({\mathbb L})$
is open and $\eta(x) = 0$ as meaning that it is closed, and represent
each realization of the process by the collection
$\omega = \{ x \in V({\mathbb L}) : \eta(x) = 1 \}$ of open sites.
We will often call $\omega$ a configuration, making no distinction
between $\omega$ and $\eta$, and will denote by $\Omega$ the set
of all $\omega$'s.
We call a path {\bf open} (resp., {\bf closed}) if all sites on the
path are open (resp., closed).
We say that an open path of $\eta$ is contained in $\omega$ since
all its sites are in $\omega$.

Let $B(r) = \{ u \in {\mathbb R}^2 : |u| \leq r \}$, where $|\cdot|$
is the Euclidean norm.
Following~\cite{ag}, we call $\phi_o: \Omega \to \Omega$ an
{\bf enhancement function} if, for each $\omega$, it satisfies
the following (locality) properties:
\begin{itemize}
\item $\phi_o(\omega)$ depends only on the restriction of
$\omega$ to ${\mathbb B}_o = V({\mathbb L}) \cap B(R_0)$,
for some fixed $R_0<\infty$,
\item $\phi_o(\omega) \subset {\mathbb B}_o$.
\end{itemize}
$R_0$ is called {\bf enhancement range}.
We extend $\phi_o$ by translations to a family
$\phi = \{ \phi_x : x \in V({\mathbb L}) \}$ of functions associated
with the lattice sites: $\phi_x(\omega) = x + \phi_o(\tau_x \omega)$,
where $\tau_x$ is the shift operator on $\Omega$ given by
$\tau_x \omega(y) = \omega(y + x)$.
We shall also consider a collection $\alpha = \{ \alpha(x) :
x \in V({\mathbb L}) \}$ of i.i.d. random variables independent
of $\eta$, taking values in $\{ 0, 1 \}$, and interpret the
value $\alpha(x) = 1$ as meaning that the enhancement at site
$x$ is ``activated.''
We denote by $s=P'_s(\alpha(x)=1)$ the {\bf density of enhancement},
where $P'_s$ is a Bernoulli product measure independent of $P_p$.
The {\bf enhanced configuration} $\hat\omega=\hat\omega(\omega,\alpha)$
is then defined as
\begin{equation}
\hat\omega = \omega \cup \left( \bigcup_{x : \alpha(x) = 1}
\phi_x(\omega)  \right)
\end{equation}
and the corresponding $\hat\eta$ is obtained by declaring open,
regardless of their state in $\eta$, all the sites contained in
$\phi_x(\omega)$ for each $x$ such that $\alpha(x)=1$.

Let us now make the concept of essential enhancement precise.
We say that a path is {\bf self-repelling}
(see~p.~66~of~\cite{grimmett}) if none of its sites is adjacent
to any other site of the path except for its two neighbors in the
path (one neighbor, in the special case of the first and last site
of the path).
An enhancement is called {\bf essential} if there exists
a configuration $\omega$ containing no doubly-infinite,
self-repelling path, but such that the enhancement at the
origin produces such a path.

Notice that for every path $\gamma$ between $x$ and $y$, there
always exists at least one self-repelling subpath between the
same sites.
To see this, take a shortest (in terms of number of sites)
subpath $\gamma'$ of the original path that connects $x$ and $y$.
If that were not a self-repelling path, it would contain at least
two adjacent sites which are not neighbors in the path.
By deleting from $\gamma'$ all the sites and edges between those
two sites (in the order inherited from the original path), one
would produce a subpath of $\gamma$ that is shorter than $\gamma'$,
thus getting a contradiction.

Like for a path, we say that a loop is self-repelling if none of its
sites is adjacent to any other site of the loop, except for its two
neighbors in the loop.
If the interior of a loop $\lambda$ is not empty and
$x \in \text{int}(\lambda)$, then there exists at least one
self-repelling subloop whose interior also contains $x$.
The proof of this fact proceeds just like the one outlined
above for a path.

We denote by $\theta(p)$ the probability that the origin belongs to an
infinite open cluster in the original process, and by $\theta(p,s)$
the corresponding probability in the (stochastically) enhanced process.
Clearly, $\theta(p,0) = \theta(p)$ and $\theta(p,s)$ is a monotonic
function of $s$.
Notice however that, while $\theta(p)$ is a monotonic function of $p$,
this is not necessarily true of $\theta(p,s)$.
The monotonicity of $\theta(p,s)$ in $p$ depends on the nature of
the enhancement function $\phi_o$.
Let $\leq$ denote the natural partial order on the set
$\{ 0, 1 \}^{V({\mathbb L})}$, which corresponds to the partial
order induced on the set $\Omega$ by the notion of inclusion.
We call the enhancement function $\phi_o$ {\bf monotonic}
(see p.~64 of~\cite{grimmett}) if, for all $\eta \leq \eta'$
(or, equivalently, $\omega \subset \omega'$),
$\phi_o(\omega)$ is a subset of $\phi_o(\omega')$.
An enhancement defined by a monotonic enhancement function
is itself called monotonic.

%

Denoting by $p_c$ the critical probability of independent site percolation,
Aizenman and Grimmett~\cite{ag} prove the following property of essential
enhancements (although for definiteness they restrict attention to the case
of site percolation on the $d$-dimensional cubic lattice, with dimension
$d \geq 2$, their method is more general and applies to other lattices as
well).

\begin{theorem} \emph{\cite{ag}} \label{ag}
Suppose $p_c > 0$, and let $s > 0$.
For any essential enhancement, 
there exists a nonempty interval $(\pi(s), p_c)$ such that $\theta(p,s) > 0$
when $\pi(s) < p < p_c$.
\end{theorem}

The reader should be warned against the temptation to weaken the
condition that the enhancement be essential.
The following are two examples taken from~\cite{ag} of slightly
weaker conditions, neither of which is sufficient to guarantee
Theorem~\ref{ag} (see \cite{ag} for an example of enhancement
satisfying 1 and 2, but for which Theorem \ref{ag} does not hold).
\begin{enumerate}
\item There exists a configuration $\omega$ which contains no
infinite cluster, but for which there is an $\alpha$ such that
the enhanced configuration $\hat\omega(\omega,\alpha)$ contains
an infinite cluster.
\item There exists an enhanced configuration $\hat\omega(\omega, \alpha)$
which contains no doubly-infinite self-repelling path if $\alpha(0)=0$,
but contains such a path if $\alpha(0)=1$.
\end{enumerate}

An essential enhancement clearly satisfies condition 2, but the
converse is not generally true.
However, it is easy to see that a monotonic enhancement $\phi$ that
satisfies condition 2 is essential by considering the configuration
$\omega'=\hat\omega(\omega,\alpha)$ with $\alpha(0)=0$.
By condition 2, $\omega'$ does not contain a doubly-infinite
self-repelling path, but since $\phi$ is monotonic and
$\omega \subset \omega'$, if we start with $\omega'$ and activate
the enhancement only at the origin, a doubly-infinite self-repelling
path is produced. Therefore, we have constructed a configuration
$\omega'$ without a doubly-infinite self-repelling path, but such
that applying the enhancement at the origin produces such a path,
which shows that $\phi$ is essential.

It is an immediate consequence of Theorem~\ref{ag} that an essential
enhancement satisfies condition 1. The next lemma states that for
monotonic enhancements \emph{in two dimensions} the converse is also
true.


\begin{lemma} \label{equivalence}
Let $\phi$ be a monotonic enhancement in two dimensions.
$\phi$ is essential if and only if it satisfies condition 1.
\end{lemma}
\noindent Lemma~\ref{equivalence} implies that a monotonic
enhancement that is not essential cannot create an infinite
cluster; therefore, for all $p \leq p_c$, with probability $1$ there
is no infinite cluster in the enhanced percolation process.
We postpone the proof of Lemma~\ref{equivalence} to Section~\ref{proof-equivalence},
but point out that it immediately yields the following result.

\begin{corollary} \label{nonessential}
\noindent A monotonic nonessential enhancement in two dimensions
does not change the nature of the phase transition -- i.e., second
order or continuous -- and leaves the critical point unchanged.
\end{corollary}


Corollary~\ref{nonessential} implies that the monotonic
nonessential enhancement of a critical percolation process is still
critical, identifying a class of critical, dependent percolation
models for which it is natural to ask about critical exponents and
scaling limits.
The restriction to monotonic enhancements is very natural, as remarked
in Section~\ref{mono-enh}.
In fact, if $\phi_o$ is not monotonic, then in general $\theta(p,s)$
is not monotonic in $p$, which can create ambiguity over the correct
definition of the critical point (i.e., there can be more than one
critical point).
For example, it is easy to think of nonmonotonic, nonessential enhancements
that can produce an infinite cluster when the density of the original
percolation process is close to zero, but fail to do so when $p$ is
just below $p_c$ (consider, for instance, site percolation on the
square lattice, and an enhancement that makes a site open only if all
of its neighbors are closed).

Corollary~\ref{nonessential} and Theorem~\ref{ag} combined imply the
following result.

\begin{theorem} \label{necess_and_suffic}
Suppose that $p_c > 0$, and let $s > 0$.
For any monotonic enhancement in two dimensions, there
exists $\pi(s)<p_c$ such that $\theta(p,s) > 0$ when $p>\pi(s)$
if and only if the enhancement is essential.
\end{theorem}

\begin{remark}
In view of Theorem~\ref{necess_and_suffic}, in the theorems
of Section~\ref{preview}, one can substitute ``For every monotonic
enhancement that does not shift the critical point" with ``For every
monotonic nonessential enhancement."
\end{remark}

\subsection{A Useful Trick} \label{trick}

We next show how to construct from an enhancement
function  $\phi_o$ a new local function $\Phi_o:\Omega \to \{o,\emptyset\}$
and a special enhanced configuration which will be useful in the proofs
of some of the results.
$\Phi_o$ is defined as follows:
\begin{equation} \label{new-phi}
\Phi_o(\omega) = \left\{ \begin{array}{ll}
\{ o \} & \mbox{if $\{ o \} \subset \phi_x(\omega)$ for some $x \in {\mathbb B}_o$} \\
\{\emptyset\} & \mbox{otherwise}
                     \end{array} \right.
\end{equation}
Notice that $\Phi_o$ is an enhancement function with enhancement
range $R=2 R_0$.
As we did previously with $\phi_o$, we extend $\Phi_o$ by translations
to a family $\Phi = \{ \Phi_x : x \in V({\mathbb L}) \}$ of functions
associated with the lattice sites.

In general, $\Phi_o$ is different from $\phi_o$, but it is monotonic
whenever $\phi_o$ is, and has the following useful property.
\begin{lemma} \label{deterministic}
The deterministic enhancement with density $s=1$ defined by $\Phi$
is the same as the deterministic enhancement with density $s=1$
defined by $\phi$.
\end{lemma}

\noindent{\bf Proof.} It is enough to observe that
$\bigcup_{x \in V({\mathbb L})} \Phi_x(\omega) =
\bigcup_{x \in V({\mathbb L})} \phi_x(\omega)$. \fbox{} \\

\noindent The configuration obtained by a deterministic enhancement
with density $s=1$ will play an important role later, in some of the
proofs; we will denote it by $\tilde\omega$ (and $\tilde\eta$).

The function $\Phi_o$ is in general simpler than $\phi_o$, since
it takes values in $\{o,\emptyset\}$ and its effect on $\eta$ is
to add at most one open site at the origin.
For $\Phi_o$, being essential means that there exists a configuration
$\omega$ that does not contain the origin and does not contain a
doubly-infinite self-repelling path, and such that $\Phi_o(\omega)=\{o\}$
and $\omega \cup \{o\}$ contains a doubly-infinite self-repelling path
(which necessarily contains the origin).
The next lemma shows that, if $\phi_o$ is monotonic, in order to
decide whether it is essential or not, we may as well consider $\Phi_o$.

\begin{lemma} \label{essential}
Let $\phi_o$ be a monotonic enhancement function.
$\phi_o$ is essential if and only if $\Phi_o$ is essential.
\end{lemma}

\noindent{\bf Proof.} The fact that if $\Phi_o$ is essential
then $\phi_o$ must also be essential is true even if $\phi_o$
is not monotonic.
To see it, take a configuration $\omega$ that does not contain
the origin and does not contain a doubly-infinite self-repelling
path, and such that $\Phi_o(\omega)=o$ and $\omega \cup o$ contains
a doubly-infinite self-repelling path.
$\Phi_o(\omega)=o$ means that $o \subset \phi_x(\omega)$ for some
$x \in {\mathbb B_o}$.
Then $\tau_x\omega$ is a configuration without a doubly-infinite
self-repelling path, but such that applying the enhancement defined
by $\phi_o$ at the origin produces such a path, which shows that
$\phi_o$ is essential.

To prove the other direction of the claim, assume that $\phi_o$
is monotonic and essential and consider a configuration $\omega$
which does not contain a doubly-infinite self-repelling path,
but such that $\omega \cup \phi_o(\omega)$ contains at least one
such path.
Let $S$ be the set of sites contained in $\phi_o(\omega)$ that
belong to the doubly-infinite self-repelling path(s) of
$\omega \cup \phi_o(\omega)$; we enumerate them in some
deterministic way and denote them by $S=\{ x_1, \ldots x_k \}$.
We now define a new configuration
$\omega'=\omega \cup \left( \bigcup_{i=1}^{k'-1} \{x_i\} \right)$
as the unique configuration that does not contain a doubly-infinite
self-repelling path but such that $\omega' \cup \{x_{k'}\}$ contains
such a path.
The monotonicity of $\phi_o$ implies that $\{x_{k'}\} \subset \phi_o(\omega')$,
from which it is easy to see that
$\Phi_o(\tau_{x_{k'}}\omega')=\{o\}$ and $\tau_{x_{k'}}\omega' \cup \{o\}$
contains a doubly-infinite self-repelling path, which shows that
$\Phi_o$ is essential. \fbox{}

\subsection{Two Simple Examples}

We give here two simple examples of monotonic nonessential enhancements,
to show that they do exist.
1)~On the square lattice, consider an enhancement that when activated
at the origin makes it open if its neighbors to the ``north," ``east"
and ``west" are open.
2)~On the triangular lattice, consider an enhancement that when activated
at the origin makes it open if at least $m$ of its neighbors are open.

The first enhancement is essential for site percolation on the
square lattice, as can be easily seen by taking a configuration
whose only open sites are the ``north," ``east" and ``west"
neigbhors of the origin and the sites in two non-adjacent,
self-repelling paths starting from the neighbors to the ``east" and
to the ``west" of the origin. Activating the enhancement at the
origin would join the two paths into a doubly-infinite open path.
However, the same enhancement is nonessential for $*$-percolation
(i.e., site percolation on the close-packed version of the square
lattice) since the ``north," ``east" and ``west" neighbors of the
origin already form a $*$-connected set, so that making the origin
open does not ``enhance" the connectivity. By a similar reasoning,
one can easily see that the second enhancement is essential if $m
\leq 4$, but nonessential if $m=5$ or $6$.

As the two examples show, whether an enhancement is essential or not depends
on the geometry of the lattice. Essential enhancements are able to target
special locations where the addition of an open site has a significant effect
on the connectivity of the clusters.

%
%

\section{More Universality Results} \label{universality}

In the following sections, we will consider three different (but
closely related) aspects of universality.
Two of them concern the continuum scaling limit, which is obtained
by considering the percolation model realized on the lattice
$\delta{\mathbb L}$ and letting the mesh $\delta$ of the rescaled
lattice go to $0$.
In the scaling limit, the range of the enhancement also gets scaled
by a factor $\delta$ (i.e., the enhancement range becomes $\delta R_0$).

We note that, although the results of the next sections are stated
for stochastically activated enhancements, it will be clear from the
proofs that they are also valid for deterministically activated
enhancements.

\subsection{Critical Exponents} \label{univ-critexp}

Consider independent (site) percolation and enhancement percolation
on $\mathbb L$. (We remind the reader that $\mathbb L$ is either the
square, triangular, or hexagonal lattice.) The following holds.

\begin{lemma} \label{mainthm-critexp}
For every monotonic nonessential enhancement, there
exist constants $0 < c_1, c_2 < \infty$ such that,
$\forall s \in [0,1]$ and $|x|$ large enough,
\begin{eqnarray}
\theta(p) \leq \theta(p,s) \leq c_1 \, \theta(p) \,\,\,
\text{ for } p \in (p_c, 1], \label{theta} \\
\tau_p(x) \leq \tau_{p,s}(x) \leq p^{-c_2} \, \tau_p(x), \,\,\,
\text{ for } p \in (0, p_c], \label{tau} \\
\xi(p,s) = \xi(p), \,\,\, \text{ for } p \in (0, p_c]. \label{xi}
\end{eqnarray}
\end{lemma}

Theorem~\ref{critexp-triangular} of Section~\ref{preview}, as well as
the next result, Theorem~\ref{cor-critexp}, follow immediately from
Lemma~\ref{mainthm-critexp}.

\begin{theorem} \label{cor-critexp}
Suppose that the critical exponents $\beta$, $\eta$, $\nu$ and $\delta$
exist for independent site percolation on $\mathbb L$.
Then, for every monotonic nonessential enhancement, the critical exponents
$\beta$, $\eta$, $\nu$ and $\delta$ exist also for the enhanced percolation
process and have the same numerical values as for the independent process
(before the enhancement).
\end{theorem}

The proofs of Lemma~\ref{mainthm-critexp} and Theorem~\ref{cor-critexp},
as well as of Theorem~\ref{critexp-triangular} of Section~\ref{critexp},
are given in Section~\ref{proof-critexp}.

\subsection{Crossing Probabilities} \label{crossing-prob}

The fact that we embedded the square, triangular and hexagonal lattices as regular
tessellations of the plane (see Figures~\ref{close-pack} and~\ref{duality}) means
that they can be partitioned into equal ``cells," a property that will be used in
the proof of Theorem~\ref{cross-prob} below.

We look at the percolation model $\hat\eta$ on $\delta {\mathbb L}$
and consider the scaling limit, as $\delta \to 0$, of crossing
probabilities, focusing for simplicity on the probability of an open
crossing of a rectangle aligned with the Cartesian coordinate axes.
A similar approach would work for any domain with a ``regular''
boundary, but it would imply dealing with more complex deformations
of the boundary than that needed for proving the result for a
rectangle.

Consider a finite rectangle
${\cal R} = {\cal R}(b,h) = (-b/2,b/2) \times (-h/2,h/2)
\subset {\mathbb R}^2$ centered at the origin of $\mathbb L$,
with sides of lengths $b$ and $h$ and aspect ratio $\rho = b/h$.
We say that there is an open vertical $\mathbb L$-crossing of ${\cal R}$
in $\eta$ (respectively, $\hat\eta$) if ${\cal R} \cap \delta{\mathbb L}$
contains an $\mathbb L$-path of open sites from $\eta$ (resp., $\hat\eta$)
joining the top and bottom sides of the rectangle ${\cal R}$, and
call $\varphi_{\delta}(b,h)$ (resp., $\hat\varphi_{\delta}(b,h)$)
the probability of a such an open crossing.
More precisely, there is a vertical open crossing in $\eta$
(resp., $\hat\eta$) if there is an $\mathbb L$-path
$(x_0, \text{e}_{x_0,x_1}, x_1, \ldots, x_m, \text{e}_{x_m,x_{m+1}}, x_{m+1})$
in $\mathbb L$ with $\eta(x_j)=1$ (resp., $\hat\eta(x_j)=1$) for all $j$,
with $\delta x_1, \ldots, \delta x_m$ all in $\cal R$, and with
the line segments $\overline{\delta x_0,\delta x_1}$ and
$\overline{\delta x_m,\delta x_{m+1}}$ touching respectively
the top side $[-b/2,b/2] \times \{ h/2 \}$ and the bottom side
$[-b/2,b/2] \times \{ -h/2 \}$.

It is believed that the scaling limit of crossing probabilities
for independent critical percolation in two dimensions exists, is
universal, and is given by Cardy's formula~\cite{cardy1,cardy2}.
However Cardy's formula has so far been rigorously proved only
in the case of critical site percolation on the triangular
lattice~\cite{smirnov,smirnov-long}, for which we have already
presented a result in Section~\ref{cardy+full} (see Theorem~\ref{cardy}
there).
For the case of the square and hexagonal lattice, assuming that
$\lim_{\delta \to 0} \varphi_{\delta}(b,h) = F(\rho)$, where $F$
is a continuous function, we have the following result.

\begin{theorem} \label{cross-prob}
Let $\mathbb L$ be the square or hexagonal lattice,
and assume that the scaling limit of crossing probabilities
of a rectangle $\cal R$ exists for independent critical site
percolation on $\mathbb L$
and is given by a continuous function $F$ of $\rho$.
Then, for every monotonic nonessential enhancement, the
corresponding crossing probabilities in the enhanced process
have the same scaling limit~$F$.
\end{theorem}

The proof of Theorem~\ref{cross-prob} (and of Theorem~\ref{cardy}
of Section~\ref{cardy+full}) is given in Section~\ref{proof-cardy}.

\subsection{The Full Scaling Limit} \label{universality-scaling}

\begin{figure}[!ht]
\begin{center}
\includegraphics[width=8cm]{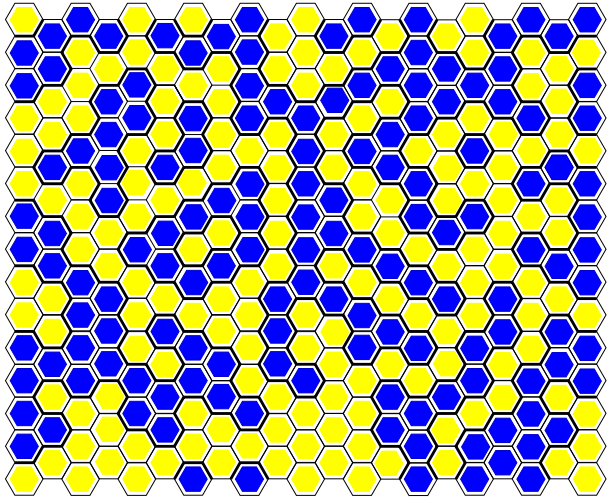}
\caption{Finite portion of a (site) percolation configuration on
the triangular lattice $\mathbb T$, where the sites of $\mathbb T$
are represented as faces of the hexagonal lattice $\mathbb H$.
The boundaries between clusters are indicated by heavy lines.}
\label{percolation}
\end{center}
\end{figure}

In this section we further restrict attention to the triangular
lattice $\mathbb T$, whose sites we think of as the (centers of the)
elementary cells of a regular hexagonal lattice $\mathbb H$ embedded
in the plane as in Figure~\ref{percolation}.
In this case, ${\mathbb L}={\mathbb L}^*={\mathbb T}$ and there is
no difference between $\mathbb L$-paths and $*$-paths.


The full scaling limit of two-dimensional critical percolation was
described in~\cite{cn2}; it represents the limit as $\delta \to 0$
of the collection of all the boundaries between open and closed
clusters at $p=p_c$.
Its existence and some of its properties have been proved in~\cite{cn3}
for the case of critical site percolation on $\mathbb T$.
In dealing with the scaling limit, as in~\cite{cn2,cn3}, we adopt the
Aizenman-Burchard approach~\cite{ab}. 
A precise formulation requires some additional notation, given below.

The edge $\text{e}^d_{x,y} = (x,y)_d$ of $\mathbb H$, dual to
the edge $(x,y)$ of $\mathbb T$ is said to be {\bf unsatisfied}
in $\eta$ (resp., in $\hat\eta$) if $\eta(x) \neq \eta(y)$
(resp., $\hat\eta(x) \neq \hat\eta(y)$).
We call $\Gamma(\eta)$ (resp., $\Gamma(\hat\eta)$) the set
of unsatisfied dual edges in the configuration $\eta$
(resp., $\hat\eta$).
The dual edges in $\Gamma$ make up the interfaces between
open and closed clusters.
More precisely, an open (resp., closed) cluster is a maximal
connected subset of $\mathbb T$ whose sites are all open (resp.,
closed), and we call {\bf boundary} between an open and a closed
cluster the collection of unsatisfied dual edges that lie between
sites of the two clusters.


A {\bf boundary path} ({\bf b-path} for short) is an
\emph{oriented}, \emph{self-avoiding} $\mathbb H$-path
$\vec\gamma_d = \{ \xi_0, \text{e}^d_1, \ldots, \text{e}^d_{k-1}, \xi_k \}$
written as an ordered, alternating sequence of sites of $\mathbb H$
and the edges between them, such that all edges in $\vec\gamma_d$
belong to $\Gamma$.
If the path $\vec\gamma_d$ forms a loop (i.e., if $\xi_k = \xi_0$),
it will be called a {\bf b-loop}.
In this case, $\vec\gamma_d$ coincides with a complete boundary.
As boundaries between open and closed clusters, b-paths can always
be extended to form a loop or a doubly-infinite path.
When there is no infinite cluster, like in two-dimensional critical
Bernoulli percolation, all complete boundaries are b-loops.


We denote by ${\cal F}_{\delta}$ a collection of b-loops of step
size $\delta$, which we identify with the boundaries of independent
(Bernoulli) percolation on the lattice $\delta {\mathbb T}$ at $p = p_c(=1/2)$,
and by $\hat{\cal F}_{\delta}$ the collection of the boundaries
obtained from ${\cal F}_{\delta}$ by enhancement.
We note that independent percolation and the enhanced one are coupled
on the probability space
$(\{ 0,1 \}^{V(\mathbb T)} \times \{ 0,1 \}^{V(\mathbb T)}, \Sigma, {\mathbb P})$,
where ${\mathbb P} = P_p \times P'_s$ and $\Sigma$ is the $\sigma$-algebra
generated by cylinder events.
We call $\mu_{\delta}$ the distribution of ${\cal F}_{\delta}$ and
$\hat\mu_{\delta}$ the distribution of $\hat{\cal F}_{\delta}$.
These collections of paths (or rather, their distributions), indexed
by $\delta$, are the objects of which we take the continuum scaling
limit, letting $\delta \to 0$.


The scaling limit $\delta \to 0$ can be taken by focusing on fixed
finite regions, $\Lambda \subset {\mathbb R}^2$, or by treating the
whole ${\mathbb R}^2$.
The second option 
avoids technical issues that arise near the boundary of $\Lambda$.
A convenient way of dealing with the whole ${\mathbb R}^2$
is to replace the Euclidean metric with a distance function
$\text{d}(\cdot,\cdot)$ defined on ${\mathbb R}^2 \times {\mathbb R}^2$
by
\begin{equation} \label{metric}
\text{d}(u,v) = \inf_f \int (1 + |f|^2)^{-1} \, dl ,
\end{equation}
where the infimum is over all smooth curves $f(l)$
joining $u$ with $v$, parametrized by arclength $l$,
and $|\cdot|$ denotes the Euclidean norm.
This metric is equivalent to the Euclidean metric in bounded regions,
but it has the advantage of making ${\mathbb R}^2$ precompact.
Adding a single point at infinity yields the compact space
$\dot{\mathbb R}^2$ which is isometric, via stereographic projection,
to the two-dimensional sphere.



Denote by ${\cal S}$ the complete, separable metric space of continuous
curves in $\dot{\mathbb R}^2$ with a distance $\text{D}(\cdot,\cdot)$
based on the metric defined by equation~(\ref{metric}) as follows.
Curves are regarded as equivalence classes of continuous functions $g(t)$,
from the unit interval $[0,1]$ to $\dot{\mathbb R}^2$, modulo monotonic
reparametrizations.
The distance D between two curves of $\cal S$, ${\cal C}_1$ and ${\cal C}_2$,
is defined by
\begin{equation}
\text{D} ({\cal C}_1,{\cal C}_2) = \inf_{f_1,f_2} \sup_{t \in [0,1]}
\text{d} (g_1 (f_1(t)), g_2 (f_2(t))),
\end{equation}
where $g_1$ and $g_2$ are particular parametrizations of
${\cal C}_1$ and ${\cal C}_2$, and the infimum is over the set of all
monotone (increasing or decreasing) continuous functions from the unit
interval onto itself.
The distance between two closed sets of curves, $\cal F$ and ${\cal F}'$,
is defined by the induced Hausdorff metric as follows:
\begin{equation} \label{hausdorff}
\left(\text{dist}({\cal F},{\cal F}') \leq \varepsilon \right)
\Leftrightarrow \left(\forall \, {\cal C} \in {\cal F}, \, \exists \,
{\cal C}' \in {\cal F}' \text{ with }
\text{D} ({\cal C},{\cal C}') \leq \varepsilon,
\text{ and vice versa} \right).
\end{equation}

With these definitions, we have the following result.

\begin{lemma} \label{main-thm}
The distance between the collection of random curves
${\cal F}_{\delta}$ from independent percolation and the corresponding
collection of curves $\hat{\cal F}_{\delta}$ from the enhanced percolation
process goes to zero almost surely as $\delta \to 0$; i.e., for
$\mathbb P$-almost every $(\eta, \alpha)$,
\begin{equation} \label{convergence}
\lim_{\delta \to 0} \text{\emph{dist}}
({\cal F}_{\delta}, \hat{\cal F}_{\delta}) = 0.
\end{equation}
\end{lemma}

The main application of Lemma~\ref{main-thm} is Theorem~\ref{full-triangular}
of Section~\ref{cardy+full}.
The proofs of Lemma~\ref{main-thm} and Theorem~\ref{full-triangular} are given
in Section~\ref{full}.

%



%

\section{Proofs of the Enhancement Percolation Results} \label{proof-equivalence}

First of all we remind the reader of the definition of
self-repelling path (see~p.~66~of~\cite{grimmett}), then we prove
Lemma~\ref{equivalence} and Theorem~\ref{necess_and_suffic} of
Section~\ref{def+main-res}. We will need three ``geometric" lemmas,
two of which are easy to establish. The third one is of the type
that can be deemed ``obvious" (at least if one focuses on a specific
lattice), but is nonetheless quite tedious to prove. The proofs, in
a general setting, of those three lemmas are given in
Appendix~\ref{matching}.

\begin{definition} \label{self-repelling}
We say that a path is {\bf self-repelling}
if none of its sites is adjacent to any other site of the path except for
its two neighbors in the path (one neighbor, in the special case of the
first and last site of the path).
\end{definition}


\noindent{\bf Proof of Lemma~\ref{equivalence}.}
As already remarked, one direction is an immediate consequence of
Theorem~\ref{ag}, so we only need to prove the other direction.
Let us consider an enhancement that satisfies condition 1 and let
$\omega$ be a configuration which does not contain an infinite cluster,
but such that an infinite cluster exists in the fully enhanced
configuration $\hat\omega(\omega,\alpha)$ for some $\alpha$.
If this is the case, then also the configuration $\tilde\omega$ defined
at the end of Section~\ref{def+main-res} contains a fortiori an infinite
cluster.
We shall also assume, without loss of generality, that $\tilde\eta(o)=1$
(i.e., $\tilde\omega$ contains the origin) and that the origin belongs
to the infinite cluster of $\tilde\omega$; then $\tilde\omega$ contains
an infinite path $\gamma$ starting at the origin.

Since $\omega$ contains no infinite cluster, $\eta$ does not contain
an infinite open cluster and the origin must be surrounded by infinitely
many closed $*$-loops.
Take one such $*$-loop that does not intersect $B(2R)$ and
construct from it a self-repelling $*$-loop $\lambda^*$ whose
interior contains $B(2R)$ (where $R=2R_0$ is the range of $\Phi_o$).
By Lemma~\ref{neighbors}, stated and proved in Appendix~\ref{matching},
each site of $\lambda^*$ has at least one neighbor in
$V({\mathbb L}) \cap \text{int}(\lambda^*)$ and one in
$V({\mathbb L}) \cap\text{ext}(\lambda^*)$. Moreover, by Lemma~\ref{partition},
whose statement and proof can be found in the same appendix,
${\mathbb L} \cap \text{int}(\lambda^*)$ is $\mathbb L$-connected.

Every $\mathbb L$-path starting at the origin and going to
infinity intersects $\lambda^*$.
Let $x_0 \in \gamma \cap \lambda^*$ be the first site
of $\lambda^*$ intersected by $\gamma$ parametrized from the
origin to infinity, and consider $\lambda^* \cap B_{x_0} (R)$,
where $B_{x_0} (R)$ denotes the open ball of radius $R$
centered at $x_0$.
Starting from $x_0$, check the sites of $\lambda^*$
in both increasing and decreasing order and, in both directions,
mark all those sites encountered before exiting $B_{x_0} (R)$
for the first time.
Call $S$ the set of sites marked in this process
and let ${\mathbb B} = V({\mathbb L}) \cap B(R)$.
$S$ partitions $B_{x_0}(R) \setminus S$
into two disjoint subsets such that no site of
the first is $\mathbb L$-adjacent to a site of
the second (see, e.g., Lemma~A.2~of~\cite{kesten}).
In the same way, its translate $\tau_{x_0} S$
partitions ${\mathbb B} \setminus \tau_{x_0} S$
in two disjoint subsets, ${\mathbb B}_1$ and ${\mathbb B}_2$,
that are not $\mathbb L$-adjacent.

Consider the following configuration $\eta'''$ and the corresponding
$\omega'''$.

\begin{equation}
\eta'''(x) = \left\{ \begin{array}{ll}
1 & \mbox{if $x \in {\mathbb B}$ and $x \notin \tau_{x_0} S$} \\
0 & \mbox{if $x \in {\mathbb B}$ and $x \in \tau_{x_0} S$} \\
0 & \mbox{if $x \notin {\mathbb B}$}
                     \end{array} \right.
\end{equation}

\noindent We now introduce a configuration $\omega' = \omega''' \cup \omega''$,
where $\omega''$ contains only two infinite, disjoint, self-repelling
$\mathbb L$-paths, $\gamma_1$ and $\gamma_2$, which are not $\mathbb L$-adjacent
and do not intersect $\tau_{x_0} S$, and such that $\gamma_1$ starts
from an $\mathbb L$-neighbor of the origin in ${\mathbb B}_1$ and does
not intersect ${\mathbb B}_2$ and $\gamma_2$ starts from an $\mathbb L$-neighbor
of the origin in ${\mathbb B}_2$ and does not intersect ${\mathbb B}_1$.
By definition, $\omega'$ contains no doubly-infinite, self-repelling path,
but such a path is produced if the origin is added.

Now notice that $\eta(x_0)=0$ and $\tilde\eta(x_0)=1$, which means
that $\Phi_{x_o}(\omega)=x_0$ or equivalently $\Phi_o(\tau_{x_0}\omega)=o$.
Since $\phi_o$ is monotonic, $\Phi_o$ is also monotonic, and since
$\eta(x+x_0) \leq \eta'(x) \, \, \, \forall x \in {\mathbb B}$,
it follows that $\Phi_o(\omega')=o$.
Therefore, activating the enhancement defined by $\Phi_o$ at the
origin in the configuration $\omega'$ produces a doubly-infinite,
self-repelling path.
The enhancement defined by $\Phi_o$ is therefore essential.
This implies, by Lemma~\ref{essential}, that also the enhancement
defined by $\phi_o$ is essential. \fbox{} \\

\noindent{\bf Proof of Theorem~\ref{necess_and_suffic}.}
One direction is already contained in Theorem~\ref{ag}
of Section~\ref{mono-enh}.
The other direction follows from Corollary~\ref{nonessential}
of Section~\ref{enhancement}, which is a straightforward
consequence of Lemma~\ref{equivalence}. \fbox{}

\section{Proofs of the Universality Results} \label{proofs}

We assume throughout this section that we are dealing with a monotonic
nonessential enhancement $\phi$ of range $R_0$. Remember that $R=2 R_0$
is the range of the enhancement $\Phi$ obtained from $\phi$, as explained
at the end of Section~\ref{def+main-res}, and that the enhancement range
gets rescaled by a factor $\delta$ when the lattice under consideration
is $\delta{\mathbb L}$.
In the rest of the paper, $B_u(r)$ will indicate the open ball of
radius $r$ centered at $u$.
We will also need the following definitions and lemmas.

\begin{definition} \label{def-protected}
A site $x \in V({\mathbb L})$ is called {\bf protected} if $\eta(x)=0$
and in $\eta$ there are at least two closed $*$-paths,
$\gamma^*_1 = \{ z^1_0, \text{\emph{e}}^1_1, z^1_1, \ldots, z^1_{k_1} \}$
and $\gamma^*_2 = \{ z^2_0, \text{\emph{e}}^2_1, z^2_1, \ldots, z^2_{k_2} \}$,
such that (i) $z^1_0 = z^2_0 = x$, (ii) $z^1_{k_1}, z^2_{k_2} \notin B_x(2R)$,
(iii) except for $x$, $\gamma^*_1$ and $\gamma^*_2$ are disjoint and nonadjacent.
\end{definition}

\begin{definition} \label{def-stable}
A dual edge $\text{\emph{e}}^d_{xy} = (x,y)_d$ such that $x$ is
protected and $\eta(y)=1$ is called {\bf stable}.
\end{definition}

Protected sites satisfy the conditions of the next lemma, which
is a key technical result.
Its proof is similar to that of Lemma~\ref{equivalence}; we spell
it out for the sake of completeness. 
We remind the reader that $\tilde\eta$ is the configuration obtained
by a deterministic enhancement with $s=1$ (see Section~\ref{trick}).

\begin{lemma} \label{lemma-stable}
Set $\delta=1$ for simplicity.
If $\eta(x)=0$ and in $\eta$ there are at least two closed $*$-paths, $\gamma^*_1$ and
$\gamma^*_2$, starting at $x$ but otherwise disjoint and nonadjacent in ${\mathbb L}^*$,
which exit $B_x(R)$, then $\tilde\eta(x)=0$.
\end{lemma}

\noindent {\bf Proof.} We will prove the result by contradiction.
Extract from $\gamma^*_1$ and $\gamma^*_2$ two self-repelling
$*$-paths, ${\gamma'}^*_1$ and ${\gamma'}^*_2$, starting at $x$, and let
$\gamma^* = {\gamma'}^*_1 \cup {\gamma'}^*_2$.
Since ${\gamma'}^*_1$ and ${\gamma'}^*_2$ are disjoint and nonadjacent in ${\mathbb L}^*$,
except for the common starting point $x$, and each one of them is self-repelling,
$\gamma^*$ is also self-repelling.
Moreover, like the set $S$ in the proof of Lemma~\ref{equivalence}, $\gamma^*$
partitions $(V({\mathbb L}) \cap B_x(R)) \setminus V(\gamma^*)$ into two disjoint,
nonadjacent, $\mathbb L$-connected sets $D_1$ and $D_2$.
From Lemma~\ref{neighbors}, site $x$ has at least one $\mathbb L$-neighbor in $D_1$
and one in $D_2$.

Consider the following configuration $\eta'''$ and the corresponding $\omega'''$.

\begin{equation}
\eta'''(y) = \left\{ \begin{array}{ll}
1 & \mbox{if $y \in {\mathbb B}$ and $y \notin \tau_x \gamma^*$} \\
0 & \mbox{if $y \in {\mathbb B}$ and $y \in \tau_x \gamma^*$} \\
0 & \mbox{if $y \notin {\mathbb B}$}
                     \end{array} \right.
\end{equation}
where ${\mathbb B} = V({\mathbb L}) \cap B(R)$.
Notice that, by Lemmas~\ref{neighbors} and \ref{partition}, $\tau_x \gamma^*$
partitions ${\mathbb B} \setminus V(\tau_x \gamma^*)$ into two disjoint, nonadjacent,
$\mathbb L$-connected sets ${\mathbb B}_1$ and ${\mathbb B}_2$, and that the
origin, contained in $\tau_x \gamma^*$, has at least one $\mathbb L$-neighbor
in ${\mathbb B}_1$ and one in ${\mathbb B}_2$.

We now introduce a new configuration $\omega' = \omega''' \cup \omega''$,
where $\omega''$ consists of only two infinite, disjoint,
self-repelling $\mathbb L$-paths, $\gamma_1$ and $\gamma_2$, that are nonadjacent
in $\mathbb L$ and do not intersect $\tau_x \gamma^*$, and such that $\gamma_1$
starts from an $\mathbb L$-neighbor of the origin in ${\mathbb B}_1$ and does
not intersect ${\mathbb B}_2$ and $\gamma_2$ starts from an $\mathbb L$-neighbor
of the origin in ${\mathbb B}_2$ and does not intersect ${\mathbb B}_1$.
By construction, $\omega'$ contains no doubly-infinite, self-repelling path,
but such a path is produced if the origin is added.

Suppose now that $\tilde\eta(x) = 1$.
Since the enhancement that produced $\tilde\eta$ is assumed
monotonic and, by construction,
$\eta(x+x_0) \leq \eta'(x) \, \, \, \forall x \in {\mathbb B}$,
$\tilde\eta(x) = 1$ implies that, when the enhancement is
activated at the origin in $\omega'$, the origin is added to
$\omega'$ and a doubly-infinite, self-repelling path is produced.
From this it follows that the enhancement is essential, giving a
contradiction. \fbox{} \\


Lemma~\ref{lemma-stable} will be used several times, starting with the
proof of another key technical result, Lemma~\ref{lemma-protected} below.

\begin{lemma} \label{lemma-protected}
Set $\delta=1$ for simplicity.
If $x$ and $y$ are protected, $B_x(2R) \cap B_y(2R) = \emptyset$, and in
$\eta$ there is a closed $*$-path $\gamma^*$ from $x$ to $y$, then in
$\tilde\eta$ there is a closed $*$-path ${\gamma'}^*$ from $x$ to $y$.
\end{lemma}

\noindent {\bf Proof.} To prove the lemma, we first construct a new closed
$*$-path ${\gamma'}^*$ joining $x$ with $y$, and then show that we can apply
Lemma~\ref{lemma-stable} to all its sites.
First of all, extract from $\gamma^*$ a self-repelling $*$-path ${\gamma''}^*$.
Part of the new path that we are going to construct will coincide with ${\gamma''}^*$;
we just need to define ${\gamma'}^*$ inside $B_x(2R)$ and $B_y(2R)$.

Let us start with $B_y(2R)$.
Remember that $y$ is protected and let $\gamma^*_{y,1}$ and $\gamma^*_{y,2}$
be the $*$-paths of Definition~\ref{def-protected}.
Call $y'$ the first site of ${\gamma''}^*$ (counting from $x$ to $y$ in the natural
order associated with ${\gamma''}^*$) that is ${\mathbb L}^*$-adjacent to a site of
either $\gamma^*_{y,1}$ or $\gamma^*_{y,2}$ (such a site always exists, although it
may coincide with an ${\mathbb L}^*$-neighbor of $y$), and assume, without loss of
generality, that $y'$ is ${\mathbb L}^*$-adjacent to $\gamma^*_{y,1}$. An analogous
construction (but with the ordering of the sites in ${\gamma''}^*$ reversed, from
$y$ to $x$) gives a site $x' \in B_x(2R)$, and again we assume, without loss of
generality, that $x'$ is ${\mathbb L}^*$-adjacent to $\gamma^*_{x,1}$.

The path ${\gamma'}^*$ is then obtained by pasting together the portion of $\gamma^*_{x,1}$
between $x$ and $x'$, the portion of ${\gamma''}^*$ between $x'$ and $y'$, and the
portion of $\gamma^*_{y,1}$ between $y'$ and $y$ in such a way that the resulting
path is self-repelling.
The sites in the new path need to be reordered, which can be easily done
starting from $x$ and using the order that each piece inherits from the
original path it comes from (or that order inverted, for the last piece).

It is now easy to see that $\tilde\eta(z) = 0$ for all $z \in {\gamma'}^*$ by an
application of Lemma~\ref{lemma-stable}.
In order to do that, we just need to check that, for each site $z \in {\gamma'}^*$,
there are two closed $*$-paths that start at $z$, exit $B_z(R)$, and are not
${\mathbb L}^*$-adjacent, except for their common starting point $z$.
If $z \in {\gamma'}^*$, but $z \notin B_x(R), B_y(R)$, this is obvious.
It suffices to take the two portions of ${\gamma'}^*$ from $z$ to $x$ and from $z$
to $y$ until they exit $B_z(R)$.

If $z \in {\gamma'}^*$ and, say, $z \in B_y(R)$, we construct the two paths in the
following way.
One path will be the portion of ${\gamma'}^*$ that from $z$ exits $B_z(R)$ in the
direction of $x$.
The other will be the portion of ${\gamma'}^*$ from $z$ to $y$ pasted together with
$\gamma^*_{y,2}$.
This last path is, by definition, not ${\mathbb L}^*$-adjacent to $\gamma^*_{y,1}$,
and cannot be ${\mathbb L}^*$-adjacent to the portion of ${\gamma'}^*$ that coincides
with ${\gamma''}^*$ because of our assumption, in choosing $y'$ to construct
${\gamma'}^*$, that this ``touches" $\gamma^*_{y,1}$ before $\gamma^*_{y,2}$. \fbox{}

\begin{remark} \label{remark-stable-protected}
Because of the monotonicity in the density of enhancement $s$,
Lemmas~\ref{lemma-stable} and~\ref{lemma-protected} immediately imply
the same conclusions for any stochastically enhanced configuration $\hat\eta$.
\end{remark}

\subsection{The Critical Exponents} \label{proof-critexp}

\noindent {\bf Proof of Lemma~\ref{mainthm-critexp}.}
The lower bound for $\theta(p,s)$ in equation~(\ref{theta}) is obvious by
monotonicity in $s$.
For the upper bound, we rely on the following observation.
Choose a positive constant $K$ so that the annulus $B(R+K) \setminus B(R)$
contains at least one $*$-loop.
If no site in $B(R+K)$ is connected to infinity by an open $\mathbb L$-path
before the enhancement takes place, then $B(R)$ must be surrounded by a closed,
self-repelling $*$-loop $\lambda^*$ (i.e., $B(R) \subset \text{int}(\lambda^*)$).
It then follows that each site in $\lambda^*$ satisfies the hypotheses of
Lemma~\ref{lemma-stable}.
Therefore, the origin will not be connected to infinity by an open
$\mathbb L$-path after the enhancement (see Remark~\ref{remark-stable-protected}).

For two subsets $C$ and $D$ of $\mathbb L$, we indicate with $\{C \longleftrightarrow D\}$
the event that some site in $C$ is connected to some site in $D$ by an open
$\mathbb L$-path, with $\{C \longleftrightarrow \infty\}$ the event that
some site in $C$ belongs to an infinite, open $\mathbb L$-path.
With this notation, we can write
\begin{equation} \label{upper-theta}
\theta(p,s) \leq P_p(B(R+K) \longleftrightarrow \infty).
\end{equation}

Since $\{ o \longleftrightarrow \infty \} \supset
\{ y \text{ is open } \forall y \in B(R+K) \} \cap \{ B(R+K) \longleftrightarrow \infty \}$,
using the FKG inequality we have
\begin{equation}
P_p(o \longleftrightarrow \infty) \geq p^c \, P_p(B(R+K) \longleftrightarrow \infty),
\end{equation}
with $c = || B(R+K) \cap {\mathbb L} ||$.
From this and (\ref{upper-theta}), we get
\begin{equation}
\theta(p,s) \leq c_1 \, \theta(p),
\end{equation}
with $c_1 = (1/p_c)^c$.

The lower bound for $\tau_{p,s}(x)$ in equation~(\ref{tau}) is again obvious.
To obtain the upper bound, we consider a site $x$ at distance larger than
$2(R+K)$ from the origin (where $K$ is the constant introduced above).
Unless $\{B(R+K) \longleftrightarrow B_x(R+K)\}$ before the enhancement takes
place, $B(R)$ and $B_x(R)$ must be separated by a closed $*$-loop surrounding
one of them or by a doubly-infinite, closed $*$-path.
Therefore, by an application of Lemma~\ref{lemma-stable}, as before, it cannot
be the case that $\{o \longleftrightarrow x\}$ after the enhancement.
This yields
\begin{equation} \label{upper-tau}
\tau_{p,s}(x) \leq P_p(B(R+K) \longleftrightarrow B_x(R+K)).
\end{equation}

Since $\{o \longleftrightarrow x\} \supset
\{ y \text{ is open } \forall y \in B(R+K) \} \cap \{ z \text{ is open } \forall z \in B_x(R+K) \}
\cap \{ B(R+K) \longleftrightarrow B_x(R+K) \}$,
using the FKG inequality we have
\begin{equation}
P_p(o \longleftrightarrow x) \geq p^{c_2} \, P_p(B(R+K) \longleftrightarrow B_x(R+K)),
\end{equation}
where $c_2 = 2 c$ is independent of $x$ and $p$.
From this and (\ref{upper-tau}), we get
\begin{equation}
\tau_{p,s}(x) \leq p^{-c_2} \, \tau_p(x),
\end{equation}
as required.

Equation~(\ref{xi}) is an immediate consequence of equation~(\ref{tau});
it is enough to observe that
\begin{equation}
\lim_{|x| \to \infty} \left\{ - \frac{1}{|x|} \left[\log \tau_p(x) - c_2 \log p \right] \right\}
= \xi(p)^{-1}. \,\,\,\,\, \fbox{}
\end{equation}

\bigskip

\noindent {\bf Proof of Theorems~\ref{critexp-triangular} and~\ref{cor-critexp}.}
It follows from (\ref{theta}) and (\ref{tau}) that
\begin{eqnarray}
- \frac{\log \theta(p)}{\log (p - p_c)} \leq - \frac{\log \theta(p,s)}{\log (p - p_c)}
\leq - \frac{\log c_1 + \log \theta(p)}{\log (p - p_c)}, \,\,\,
\text{ for } p \in (p_c,1], \\
\frac{\log \tau_{p_c}(x)}{\log |x|} \leq \frac{\log \tau_{p_c,s}(x)}{\log |x|}
\leq \frac{\log \tau_{p_c}(x) - c_2 \log p}{\log |x|}, \,\,\,
\text{ for } |x| \text{ large enough}.
\end{eqnarray}
For $p \in (0,p_c)$, observing that
$\chi(p) = E_p \sum_{x \in {\mathbb L}} I_{\{o \longleftrightarrow x\}}
= \sum_{x \in {\mathbb L}} \tau_p(x)$ (where $I_{\{\cdot\}}$ is the indicator
function), (\ref{tau}) yields $\chi(p) \leq \chi(p,s) \leq p^{-c_2} \chi(p)$,
and therefore
\begin{equation}
- \frac{\log \chi(p)}{\log (p - p_c)} \leq - \frac{\log \chi(p,s)}{\log (p - p_c)}
\leq - \frac{\log \chi(p) - c_2 \log p}{\log (p - p_c)}.
\end{equation}
Using these three equations, together with equation~(\ref{xi}) and the definitions
of the critical exponents, and taking the appropriate limits concludes the proof. \fbox{}

\subsection{Crossing Probabilities} \label{proof-cardy}

{\bf Proof of Theorems~\ref{cardy} and~\ref{cross-prob}.}
We begin with a definition that will be useful in the proof.
For $(x,x')$ an ordered pair of $\mathbb L$-neighbors, we define
the {\bf partial cluster} $C_{(x,x')}$ to be the set of sites
$y \in \mathbb L$ such that there is an $\mathbb L$-path
$(x_0=x',\text{e}_{x_0,x_1},x_1, \dots, x_k=y)$ with $x_1 \neq x$
whose sites are all open or all closed.

To prove the theorem we need to compare crossing probabilities
in $\hat\eta$ with crossing probabilities in $\eta$.
In order to do that, we will use the natural coupling that
exists between $\eta$ and $\hat\eta$ via the enhancement.
First of all notice that, if an open vertical crossing of
$\cal R$ is present in $\eta$, it is also present in $\hat\eta$.
Therefore, recalling that $\varphi_{\delta}(b,h)$ (resp.,
$\hat\varphi_{\delta}(b,h)$) is the probability of an open vertical
$\mathbb L$-crossing of $\cal R$ from $\eta$ (resp., $\hat\eta$),
we have
\begin{equation} \label{lower-bound}
\liminf_{\delta \to 0} \hat\varphi_{\delta}(b,h) \geq
\lim_{\delta \to 0} \varphi_{\delta}(b,h) = F(\rho).
\end{equation}

On the other hand, if an open vertical $\mathbb L$-crossing of
$\cal R$ is not present in $\eta$, this implies the existence
of a closed horizontal ${\mathbb L}^*$-crossing of $\cal R$.
For $\delta$ small, such a crossing must involve many sites,
and the probability of finding ``near'' its endpoints two sites,
$x$ and $y$, belonging to the crossing and attached through closed
paths to two protected sites, $x'$ and $y'$, must be close to one.
If such protected sites are found, it follows from Lemma~\ref{lemma-protected}
(and Remark~\ref{remark-stable-protected}) that, when $\delta$ is
small enough, at least the portion of the closed horizontal crossing
from $x$ to $y$ is still present in $\hat\eta$.
This suggests that, conditioned on having in $\eta$ a closed
horizontal $*$-crossing of a slightly bigger (in the horizontal
direction) rectangle, with high probability there will be
in $\hat\eta$ a closed horizontal $*$-crossing of $\cal R$
blocking any open vertical $\mathbb L$-crossing.
It is then enough to prove that this probability
goes to one as $\delta \to 0$.

We will now make this more precise, adapting the
proof of Theorem~1 of~\cite{cns1}.
Consider the rectangle ${\cal R}'= {\cal R}(b',h)$ with $b'$
slightly larger than $b$ and aspect ratio $\rho'=b'/h$, and
let $\psi^*_{\delta}(b',h)$ be the probability of a closed
horizontal $*$-crossing of ${\cal R}(b',h)$ from $\eta$.
The presence of a closed horizontal $*$-crossing of ${\cal R}(b',h)$
prevents any open vertical $\mathbb L$-crossing of ${\cal R}(b',h)$
and vice versa.
Therefore,
\begin{equation} \label{inequality}
\lim_{\delta \to 0} \varphi_{\delta} (b,h) \leq
\lim_{\delta \to 0} ( 1 - \psi^*_{\delta} (b',h) )
= F(b'/h).
\end{equation}
Since, by continuity,
\begin{equation} \label{lim2}
\lim_{b' \to b} F(b'/h) = F(\rho),
\end{equation}
if we could replace $\lim_{\delta \to 0} \varphi_{\delta}(b,h)$
with $\limsup_{\delta \to 0} \hat\varphi_{\delta}(b,h)$
in~(\ref{inequality}), we would be done.

This is achieved by showing that if there is a closed
horizontal $*$-crossing $\gamma^*=(y_0, \dots, y_k)$ of
${\cal R}'$ in $\eta$, then there is a closed horizontal
$*$-crossing of ${\cal R}$ in $\hat\eta$ with probability
going to one as $\delta \to 0$.

To do this, we take a $b''$ between $b$ and $b'$ and consider
the rectangle ${\cal R}(b'',h)$.
Assume that $\gamma^*$ is parametrized from left to right
and let $y_{k_1+1}$ be the first site of $\gamma^*$ outside
of ${\cal R}'' \setminus {\cal R}$.
Analogously, let $y_{k_2-1}$ be the first site of $\gamma^*$
outside of ${\cal R}'' \setminus {\cal R}$ when $\gamma^*$
is parametrized in reversed order, from right to left.
If we can find two protected sites, $x_1$ and $x_2$, one in
each of the partial clusters $C_{(y_{k_1+1},y_{k_1})}$ and
$C_{(y_{k_2-1},y_{k_2})}$, contained inside
${\cal R}' \setminus {\cal R}''$, then we can use
Lemma~\ref{lemma-protected} to conclude that there
is a closed $*$-path from $x_1$ to $x_2$ in $\hat\eta$.
It also follows from the proof of the lemma that, for $\delta$
small enough compared to $b''-b'$, that path contains a subpath
of the portion of $\gamma^*$ inside $\cal R$, providing a closed
horizontal crossing of $\cal R$ in $\hat\eta$.

Let $A$ be the event that there is no protected site in either
the portion of $C_{(y_{k_1+1},y_{k_1})}$ contained inside
${\cal R}' \setminus {\cal R}''$ or the portion of
$C_{(y_{k_2-1},y_{k_2})}$ contained inside ${\cal R}' \setminus {\cal R}''$.
To conclude the proof, we need to show that the probability of $A$ goes
to zero as $\delta \to 0$.
To do so, we first partition ${\mathbb L}={\mathbb Z}^2,{\mathbb T}$
or $\mathbb H$ into disjoint regions $Q_i$, as explained below, and
denote by $\cal Q$ the collection of these regions.
We consider specific embeddings for the three lattices, corresponding
to the only three possible regular tessellations of the plane (see, e.g.,
\cite{bc}).
In the corresponding embeddings, the square lattice can be partitioned
into squares, and the hexagonal and triangular lattices into hexagonal
regions (this can be maybe better understood by looking at the dual lattices
and interpreting each face as a site of the original lattice -- see, for
example, Figure~\ref{bootstrap_fig3} for how to partition the triangular
lattice using the hexagonal one).
The regions $Q_i$ must be chosen large enough (depending on $R$)
so that there is positive probability for a region to contain at
least one protected site.

We now do an algorithmic construction (a related algorithmic
construction is described in~\cite{fn}) of the portions of the
partial clusters $C_{(y_{k_1+1},y_{k_1})}$ and
$C_{(y_{k_2-1},y_{k_2})}$ contained in ${\cal R}'\setminus {\cal
R}''$. We describe briefly how to do this for
$C_{(y_{k_1+1},y_{k_1})}$ (the construction for
$C_{(y_{k_2-1},y_{k_2})}$ is the same). The idea is that one starts
by setting $C_{(y_{k_1+1},y_{k_1})}$ equal to $y_{k_1}$, then looks
at $y_{k_1}$'s neighbors contained in ${\cal R}'\setminus {\cal
R}''$, and adds to $C_{(y_{k_1+1},y_{k_1})}$ those neighbors that
are closed. Then one proceeds by looking at $y_{k_1}$'s next-nearest
neighbors contained in ${\cal R}'\setminus {\cal R}''$, and so on.
This, however, is done in such a way that, when the first site in a
region $Q_i$ from $\cal Q$ is checked and found to be closed, then
the other sites in that region are checked next (in some
deterministic order), before moving to a neighboring region. The
construction stops when all of the portion of
$C_{(y_{k_1+1},y_{k_1})}$ contained in ${\cal R}'\setminus {\cal
R}''$ has been ``discovered" (together with its boundary of open
sites).

Notice that the portions of the partial clusters $C_{(y_{k_1+1},y_{k_1})}$
and $C_{(y_{k_2-1},y_{k_2})}$ inside ${\cal R}' \setminus {\cal R}''$
contain at least a number of sites of the order of $(b''-b')/\delta$,
since they contain the paths $(y_0, \dots, y_{k_1})$ and $(y_{k_2},\dots,y_k)$
respectively.
Therefore, if $b',b''$ and the size of the $Q_i$'s are fixed, since each
$Q_i$ contains a protected site with positive probability, the algorithmic
construction shows that $P_{p_c}(A) \leq \exp{(- c (b'-b'')/\delta)}$, for
some $c>0$. \fbox \\

\subsection{The Full Scaling Limit} \label{full}




To begin with, we need two preliminary results, the first one of
which is a consequence of the fact that before the enhancement we
are dealing with a Bernoulli product measure $P_p$ and is valid for
all $p \in (0,1)$.
In the next lemma (and elsewhere), the {\bf diameter} $\text{diam}(\cdot)$
of a subset of ${\mathbb R}^2$ is defined as the maximal Euclidean
distance between any two points of that subset.

\begin{lemma} \label{exp-bound}
Let $(x,y)_d$ be any (deterministic) dual edge; then
for $M$ large enough and for some constant $c>0$,
\begin{equation}
P_p \left( \exists \, \vec\gamma_d \ni (x,y)_d :
\text{\emph{diam}} (\vec\gamma_d) \geq M \text{ and } \vec\gamma_d
\text{ does not contain a stable edge} \right)
\leq e^{- c M}.
\end{equation}
\end{lemma}

\begin{figure}[!ht]
\begin{center}
\includegraphics[width=8cm]{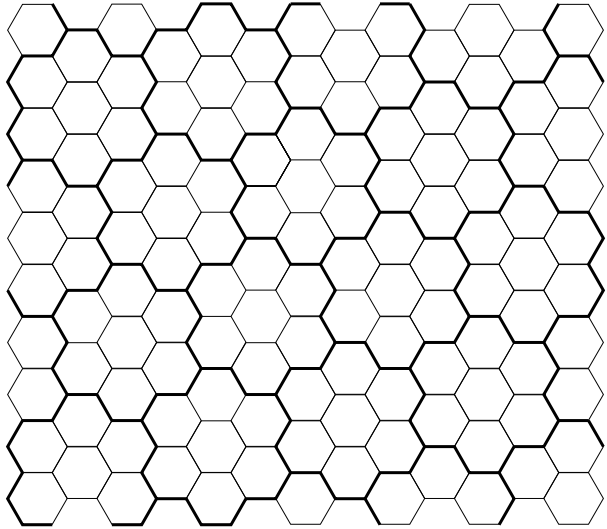}
\caption{Partition of the hexagonal lattice into
hexagonal cells.
Each hexagon represents a site of $\mathbb T$.
The process can be repeated iteratively, using
the new cells instead of the original hexagons,
to obtain a partition made of larger cells.}
\label{bootstrap_fig3}
\end{center}
\end{figure}

\noindent {\bf Proof}.
The proof requires partitioning the lattice $\mathbb H$, dual of $\mathbb T$,
into identical regions $Q_i$ and performing an algorithmic construction of
$\vec\gamma_d$, starting from $(x,y)_d$, as a percolation exploration process,
but with the additional rule that, when the exploration process enters a $Q_i$
for the first time, all the sites of $\mathbb T$ inside $Q_i$ are checked next,
according to some deterministic order.
The regions $Q_i$ can be constructed iteratively as explained in Figure~\ref{bootstrap_fig3};
their size will depend on the range $R$ of the enhancement, and has to be chosen
large enough so that, whenever the exploration process enters an unexplored region
$Q_i$, there is a strictly positive probability, bounded away from zero by a constant
that does not depend on the past history of the exploration process, that a stable
(dual) edge belonging to $\vec\gamma_d$ is found inside $Q_i$.
Let $F_i$ denote such an event.

To see that the probability of $F_i$ can be bounded below by a positive constant,
notice first that, because of the geometry of the $Q_i$'s, from every entrance
point of a $Q_i$, there is a choice of the values (open or closed) of the sites of
$\mathbb T$ in the outermost layer of $Q_i$ that forces the exploration process
to enter that region, regardless of the past history of the exploration process.
Moreover, each new region that the exploration process enters is ``virgin'' territory,
on which no information is available.

If $K$ is the number of dual edges contained in each region $Q_i$, then clearly
the exploration process has to visit at least $M/K$ regions $Q_i$, at each
new visit having a chance of ``bumping into'' a stable edge.
The bound in Lemma~\ref{exp-bound} follows immediately when $M$ is large enough
compared to $K$. \fbox{} \\

At criticality, any b-path is almost surely part of a complete
boundary which forms a b-loop.
The next lemma identifies a large ``ancestor''
$\vec\gamma_d' \subset \Gamma(\eta)$ for each large enough
b-loop $\vec\gamma_d \subset \Gamma(\hat\eta)$.

\begin{lemma} \label{ancestor}
Set $\delta=1$ for simplicity; then there is a one to one mapping from
b-loops $\vec\gamma_d \in \Gamma(\hat\eta)$ with
$\text{\emph{diam}}(\vec\gamma_d) \geq 6R$ to parent b-loops
$\vec\gamma_d' \in \Gamma(\eta)$ such that
$\text{\emph{diam}}(\vec\gamma_d') \geq \text{\emph{diam}}(\vec\gamma_d) - 4R$.
\end{lemma}

\noindent {\bf Proof.} First of all we remind the reader that in this
case ${\mathbb L}={\mathbb L}^*={\mathbb T}$, so that $\mathbb L$-paths
and $*$-paths are the same.
If the b-loop $\vec\gamma_d$ is the external boundary of a closed
cluster, such a cluster must come from a parent cluster in $\eta$ whose
diameter is equal to or bigger than that of the cluster in $\hat \eta$,
since a closed cluster can only shrink.
The external boundary $\vec\gamma_d'$ of the closed cluster from
$\eta$ is then taken to be $\vec\gamma_d$'s parent b-loop.
To establish the one to one correspondence in this case, we have to show
that two b-loops from $\hat\eta$ of diameter at least $6R$ cannot have
the same parent b-loop in $\eta$.
This can only happen if a closed cluster $C$ from $\eta$ of diameter
larger than $6R$ splits in $\hat\eta$ into two closed clusters, $C_1$
and $C_2$, each of diameter larger than $6R$.
This implies that we can find a $\mathbb T$-path $\gamma_0$, joining $C_1$
with $C_2$ and completely contained in $C$, such that, $\forall x \in \gamma_0$,
$\eta(x)=0$ and $\hat\eta(x)=1$.
However, $\gamma_0$ can be extended by two disjoint, nonadjacent
$\mathbb T$-paths, $\gamma_1$ and $\gamma_2$, of diameter larger than
$R$, with $\gamma_1$ contained in $C_1$ and $\gamma_2$ contained in $C_2$.
This shows that each site $x \in \gamma_0$ satisfies the conditions of
Lemma~\ref{lemma-stable}, so that $\hat\eta(x)=0$, leading to a contradiction.

If the b-loop $\vec\gamma_d$ is the external boundary of an open cluster $C$
from $\hat\eta$, consider two portions, $\vec\gamma_{d,1}$ and $\vec\gamma_{d,2}$,
of $\vec\gamma_d$ such that
$R \leq \text{diam}(\vec\gamma_{d,1}), \text{diam}(\vec\gamma_{d,2}) \leq 2 R$ and
$\sup_{u \in \vec\gamma_{d,1}, v \in \vec\gamma_{d,2}} |u-v| = \text{diam}(\vec\gamma_d)$
(so that $\inf_{u \in \vec\gamma_{d,1}, v \in \vec\gamma_{d,2}} |u-v| \geq
\text{diam}(\vec\gamma_d) - 4 R$).
Call $\vec\gamma_{d,3}$ and $\vec\gamma_{d,4}$ the two remaining portions
of $\vec\gamma_d$.
In $\eta$, one of the two following statements must be true: either
(1) $\vec\gamma_{d,1}$ and $\vec\gamma_{d,2}$ are connected by an open
$\mathbb T$-path, or
(2) $\vec\gamma_{d,3}$ and $\vec\gamma_{d,4}$ are connected by a closed
$\mathbb T$-path.
If (1) happens, then the path connecting $\vec\gamma_{d,1}$ and $\vec\gamma_{d,2}$
is the ``backbone'' of an open cluster whose external boundary we take as
$\vec\gamma'_d$.
The inequality $\text{diam}(\vec\gamma_d') \geq \text{diam}(\vec\gamma_d) - 4R$
is clearly satisfied.
The uniqueness of $\vec\gamma'_d$ comes from the fact that two open clusters
from $\eta$ of diameter at least $2R$ cannot merge, since each one of them
is surrounded by a closed, self-repelling $\mathbb T$-loop (of diameter
at leat $2R$) such that each site of the loop satisfies the conditions
in Lemma~\ref{lemma-stable} and is therefore closed in $\hat\eta$.

To conclude the proof, we show by contradiction that (2) cannot happen.
The open cluster $C$ of which $\vec\gamma_d$ is the external boundary
is surrounded (in $\hat\eta$) by a closed $\mathbb T$-loop $\lambda$,
with every dual edge in $\vec\gamma_d$ separating a site of $C$ from
one in $\lambda$.
Call $\lambda_3$ and $\lambda_4$ the portions of $\lambda$ corresponding
respectively to $\vec\gamma_{d,3}$ and $\vec\gamma_{d,4}$, in the sense
that the sites of $\lambda_3$ and $\lambda_4$ are next to edges of
$\vec\gamma_{d,3}$ and $\vec\gamma_{d,4}$.
Notice that $\lambda_3$ and $\lambda_4$ are closed (in $\hat\eta$
and therefore also in $\eta$) $\mathbb T$-paths and that they are
not $\mathbb T$-adjacent, since otherwise the set $C$ would not be
$\mathbb T$-connected.

Suppose that (2) did happen, then there would be a $\mathbb T$-path $\gamma'$,
with $\eta(x) = 0 \,\, \forall x \in \gamma'$, contained in
$\text{int}(\vec\gamma_d)$ and connecting $\lambda_3$ with $\lambda_4$.
From $\gamma'$ one can extract a self-repelling $\mathbb T$-path $\gamma$
joining $\lambda_3$ with $\lambda_4$ and with the property that it is not
$\mathbb T$-adjacent to $\lambda_3$ or $\lambda_4$ if one excludes its
first and last sites (this can be done by taking a minimal $\mathbb T$-path
joining $\lambda_3$ with $\lambda_4$).
Every site $x \in \gamma$ is then connected by two self-repelling, closed
$\mathbb T$-paths to $\lambda_3$ and $\lambda_4$, using which the two paths
can be continued until they exit $B_x(R)$.
If we exclude $x$, the two resulting paths are disjoint and are not
$\mathbb T$-adjacent, because of the assumptions on $\vec\gamma_d$,
$\vec\gamma_{d,1}$, and $\vec\gamma_{d,2}$.
Thus, by Lemma~\ref{lemma-stable}, $\hat \eta(x)=0, \ \forall x \in \gamma$,
contradicting the assumption that $\vec\gamma_d$ is the external boundary
of an open $\mathbb T$-cluster from $\hat\eta$. \fbox{} \\

\noindent {\bf Proof of Lemma~\ref{main-thm}.}
We can finally proceed to the proof of Lemma~\ref{main-thm} itself.
Let us start, for simplicity, with the case of a single ``large" b-loop
$\vec{\gamma}_d$ from $\eta$ of diameter at least $\varepsilon/2$ for
some fixed $\varepsilon>0$.
In the case of a ``large" loop $\vec{\gamma}_d$, as $\delta \to 0$,
we can apply Lemma~\ref{ancestor} to obtain a daughter loop $\vec{\gamma}_d'$.
We then have to show that for appropriate parametrizations $g$ and $g'$
of $\vec{\gamma}_d$ and $\vec{\gamma}_d'$, and for $\delta$ small enough,
\begin{equation} \label{small-distance}
\sup_{t \in [0,1]} \text{d}(g(t), g'(t)) < \varepsilon + 2 \delta.
\end{equation}
%
Once this is done, to prove the other direction for a single curve (i.e.,
given $\vec{\gamma}_d' \in \hat{\cal F}_{\delta}$ and $g'$, we need to find
$\vec\gamma_d \in {\cal F}_{\delta}$ and $g$ so that (\ref{small-distance}),
in which the dependence on the scale factor $\delta$ has been suppressed,
is valid), we use Lemma~\ref{ancestor}, which identifies a large parent
b-loop in $\Gamma(\eta)$ for each large b-loop in $\Gamma(\hat\eta)$.
Later we will require that both directions hold simultaneously for all the loops
(``large" and ``small") in ${\cal F}_{\delta}$ and $\hat{\cal F}_{\delta}$,
as implied by (\ref{convergence}).

For a given $\varepsilon > 0$, we divide ${\mathbb R}^2$ into two regions:
$B(6/\varepsilon)$ and
${\mathbb R}^2 \setminus B(6/\varepsilon)$.
We start by showing that, thanks to the choice of the metric~(\ref{metric}),
one only has to worry about curves (or polygonal paths) that
intersect $B(6/\varepsilon)$.
Indeed, the distance between any two points $u,v \in
\dot{\mathbb R}^2 \setminus B(6/\varepsilon)$ satisfies the
following bound
\begin{equation}
\text{d}(u,v) \leq \text{d}(u,\infty) + \text{d}(v,\infty) \leq
2 \int_0^{\infty} [1 + (l + 6 / \varepsilon)^2]^{-1} \, dl < \varepsilon/3.
\end{equation}
Thus, given any curve in ${\cal F}_{\delta}$ contained completely in
$\dot{\mathbb R}^2 \setminus B(6/\varepsilon)$, it can be approximated
by any curve in $\hat{\cal F}_{\delta}$ also contained in
$\dot{\mathbb R}^2 \setminus B(6/\varepsilon)$, and vice versa.
The existence of such curves in ${\cal F}_{\delta}$ is clearly not a problem,
since the region $\dot{\mathbb R}^2 \setminus B(6/\varepsilon)$
contains an infinite subset of $\delta{\mathbb H}$ and therefore there
is zero probability that it doesn't contain any b-path in $\eta$.
There is also zero probability that it contains no stable edge in $\eta$,
but any such edge also belongs to $\hat {\cal F}_{\delta}$.
Therefore, in the rest of the proof we consider only b-paths that
intersect $B(6/\varepsilon)$.

Given a b-path
$\vec\gamma_d = \{ \xi_0, \text{e}^d_1, \ldots, \text{e}^d_k, \xi_k \}$
in ${\cal F}_{\delta}$ with parametrization $g(t)$, we let $u_0=\xi_0$.
We will indicate by $\vec\gamma_d (u,v)$, with $u,v \in \vec\gamma_d$,
the portion of $\vec\gamma_d$ between $u$ and $v$.
The following algorithmic construction produces a sequence
$u_0, \ldots, u_N$ of points in $\vec\gamma_d$.
\begin{enumerate}
\item Start with $u_0$.
\item Once $u_0, \ldots, u_i$ have been constructed,
if $u_i \in B(6/\varepsilon)$,
let $u_{i+1}$ be the first intersection of
$\vec\gamma_d \setminus \vec\gamma_d (u_0, u_i)$
with $\partial B_{u_i}(\varepsilon / 3)$,
if $u_i \notin B(6/\varepsilon)$,
let $u_{i+1}$ be the first intersection of
$\vec\gamma_d \setminus \vec\gamma_d (u_0, u_i)$
with $\partial B^{\text d}_{u_i}(\varepsilon / 3)$.
\item Terminate when there is no next $u_i$.
\end{enumerate}

The algorithm stops after a finite number, $N$, of steps.
During the construction of the sequence $u_0, \ldots, u_N$,
$\vec\gamma_d$ is split in $N+1$ pieces, the first $N$ having
diameter at least $\varepsilon / 3$.
The construction also produces a sequence of balls
$B_{u_i}(\varepsilon / 3)$ or $B^{\text d}_{u_i}(\varepsilon / 3)$,
$i = 0, \ldots, N$.
Notice that no two successive $u_i$'s can lie outside of
$B(6 / \varepsilon)$.
In fact, if for some $i$, $u_i$ lies outside of $B(6 / \varepsilon)$,
$u_{i+1}$ belongs to $\partial B^{\text d}_{u_i}(\varepsilon / 3)$,
which is contained inside $B(6 / \varepsilon)$, due to the choice
of the metric~(\ref{metric}).
Each $u_i$ lies on a site or an edge of $\delta{\mathbb H}$, but
no more than one $u_i$ can lie on the same site or edge since
$\vec\gamma_d$ is self-avoiding and cannot use the same site or
edge more than once.
Therefore the number of $u_i$'s in $B(6 / \varepsilon)$ is bounded
by a constant times the number of sites and edges contained in
$B(6 / \varepsilon)$.
Also, the number of $u_i$'s lying outside of $B(6 / \varepsilon)$
cannot be larger than (one plus) the number of the $u_i$'s lying
inside $B(6 / \varepsilon)$.
Therefore, $N \leq const \times (\varepsilon \delta)^{-2}$.

For each $i = 0, \ldots, N-1$, let $O_i = V_i \cup V_{i+1}$, where
$V_i$ is $B_{u_i}(\varepsilon/3 + \delta)$ if
$u_i \in B(6 / \varepsilon)$ and $B^{\text d}_{u_i}(\varepsilon/3 + \delta)$
if $u_i \notin B(6 / \varepsilon)$.
Now let $v_i$ be the first intersection of $\vec\gamma_d(u_i, u_{i+1})$
with $B_{u_i}(\varepsilon / 9)$ and assume that there exists a sequence
$\bar {\text e}^d_0 ,\ldots, \bar {\text e}^d_{N-1}$ of stable edges
(see Definition~\ref{def-stable}) of $\vec\gamma_d$ with
$\bar {\text e}^d_i$ contained in $\vec\gamma_d (u_i, v_i)$.
$\vec\gamma_d (\bar {\text e}^d_i, \bar {\text e}^d_{i+1})$ is contained in
$O_i$ (for fixed $\varepsilon$ and small enough $\delta$).
Besides, for fixed $\varepsilon$ and small enough $\delta$,
any two successive stable edges $\bar {\text e}^d_i, \bar {\text e}^d_{i+1}$
lie next to two protected sites that satisfy the conditions in
Lemma~\ref{lemma-protected}, where the closed $*$-path $\gamma^*$
can be taken to be a self-repelling path of closed sites along the
b-path $\vec\gamma_d (\bar {\text e}^d_i, \bar {\text e}^d_{i+1})$
and is therefore contained in $O_i$.
As can be seen from the proof of the lemma, the path ${\gamma'}^*$
constructed from $\gamma^*$ is also contained in $O_i$.
${\gamma'}^*$ represents a barrier beyond which $\vec\gamma_d$
cannot move, since all of its sites are closed in $\hat \eta$,
so that $\vec{\gamma}_d' (\bar {\text e}^d_i, \bar {\text e}^d_{i+1})$
is confined to lie within $\vec\gamma_d (\bar {\text e}^d_i, \bar {\text e}^d_{i+1})$
and ${\gamma'}^*$, and thus within $O_i$.

To parameterize $\vec{\gamma}_d'$, we use any parametrization $g'(t)$ such
that $g'(t)=g(t)$ whenever $g(t) \in \bar {\text e}^d_i$.
Using this parametrization and the previous fact, it is clear that
the distance between $\vec\gamma_d (\bar {\text e}^d_i, \bar {\text e}^d_{i+1})$
and $\vec{\gamma}_d' (\bar {\text e}^d_i, \bar {\text e}^d_{i+1})$ does not
exceed $\varepsilon + 2 \delta$.
Therefore, conditioning on the existence of the above sequence
$\bar {\text e}^d_0, \ldots, \bar {\text e}^d_{N-1}$ of stable
edges of $\vec\gamma_d$, we can conclude that
\begin{equation} \label{small-distance1}
\sup_{t \in [0,1]} \text{d}(g(t), g'(t)) < \varepsilon + 2 \delta.
\end{equation}

It remains to prove the existence of the sequence
$\bar {\text e}^d_0, \ldots, \bar {\text e}^d_{N-1}$ of stable edges.
To do that, let us call $A_i$ the event that $\vec\gamma_d (u_i, v_i)$
does \emph{not} contain at least one stable edge, and let
$A = \cup_{i=0}^{N-1} A_i$.
Then, considering that the total number of dual edges contained in
$B(6 / \varepsilon)$ is bounded above by $const \times (\varepsilon \delta)^{-2}$
and using Lemma~\ref{exp-bound}, we have
\begin{equation} \label{small-prob}
P_p(A)
\leq (\varepsilon \delta)^{-2} e^{- c' (\varepsilon / \delta)}
\end{equation}
for some $c' > 0$.
Equation~(\ref{small-prob}) means that the probability of
not finding at least one stable edge in $\vec\gamma_d(u_i,v_i)$
for each $i=0,\ldots,N-1$ is very small and goes to $0$, for
fixed $\varepsilon$, as $\delta \to 0$.
This is enough to conclude that, with high probability
(going to $1$ as $\delta \to 0$), equation~(\ref{small-distance1})
holds.

This proves one direction of the claim, in the case of a single curve.
To obtain the other direction, as already explained, we use
Lemma~\ref{ancestor} which identifies a parent b-loop in $\Gamma(\eta)$
for each large b-loop in $\Gamma(\hat\eta)$.
Lemma~\ref{ancestor} also insures that the parent b-loop is large
enough so that we can apply to it the above arguments and obtain
the desired result.

At this point, we need to show that the above argument can be
repeated and the construction done simultaneously for all curves
in ${\cal F}_{\delta}$ and $\hat{\cal F}_{\delta}$.
First of all notice that for a fixed $\varepsilon$, any b-path
$\vec\gamma_d$ of diameter less than $\varepsilon / 2$ can be
approximated by a closest stable edge, provided that one is found
within a ball of radius $\varepsilon / 2$ that contains $\vec\gamma_d$,
with the probability of this last event clearly going to $1$ as
$\delta \to 0$, when we restrict attention to $B(6 / \varepsilon)$.
For a b-path outside $B(6 / \varepsilon)$, we already noticed
that it can be approximated by any other b-path also outside
$B(6 / \varepsilon)$.
As for the remaining b-paths, notice that the total number of
boundaries that intersect the ball $B(6/\varepsilon)$
cannot exceed $const \times (\varepsilon \delta)^{-2}$.
Thus, we can carry out the above construction simultaneously
for all the boundaries that touch $B(6/\varepsilon)$, having
to deal with at most $const \times (\varepsilon \delta)^{-2}$
segments of b-paths of diameter at least $\varepsilon/2$.
Therefore, letting
$Y_{\delta} = \text{dist}({\cal F}_{\delta},\hat{\cal F}_{\delta})$,
we can apply once again Lemma~\ref{exp-bound} and conclude that
\begin{equation} \label{small-prob1}
{\mathbb P}(Y_{\delta} > \varepsilon)
\leq (\varepsilon \delta)^{-2} e^{- c'' (\varepsilon / \delta)},
\end{equation}
for some $c''>0$.

To show that $Y_{\delta} \to 0$ ${\mathbb P}$-almost surely, as $\delta \to 0$,
and thus conclude the proof, it suffices to show that, $\forall \varepsilon>0$,
${\mathbb P}(\limsup_{\delta \to 0} Y_{\delta} > \varepsilon) = 0$.
To that end, first take a sequence $\delta_k = 1/2^k$ and notice that
\begin{equation} \label{bc}
\sum_{k=0}^{\infty} {\mathbb P}(Y_{\delta_k} > \varepsilon)
\leq \sum_{k=0}^{\infty} \frac{4^k}{\varepsilon^2} e^{- c'' 2^k \varepsilon}
< \infty,
\end{equation}
where we have made use of (\ref{small-prob1}).
Equation~(\ref{bc}) implies that we can apply the Borel-Cantelli lemma
and deduce that
${\mathbb P}(\limsup_{k \to \infty} Y_{\delta_k} > \varepsilon) = 0$,
$\forall \varepsilon>0$.
In order to handle the values of $\delta$ not in the sequence $\delta_k$,
that is those $\delta$ such that $\delta_{k+1} < \delta < \delta_k$ for
some $k$, we use the double bound
\begin{equation} \label{db}
a \, \text{d}(u,v) \leq \text{d}(a u, a v)
\leq \frac{1}{a} \, \text{d}(u,v),
\end{equation}
valid for any $0< a < 1$, which implies that $a \, Y_{\delta_k} \leq Y_{a \delta_k}
\leq \frac{1}{a} \, Y_{\delta_k}$.
The two bounds in equation (\ref{db}) come from writing
$\text{d}(a u, a v)$ as
$\text{d}(a u, a v) = \inf_{f'} \int (1 + |f'|^2)^{-1} \, dl'
= a \, \inf_{f} \int (1 + a^2 |f|^2)^{-1} \, dl $,
where $f'(l')$ are smooth curves joining $a u$ with $a v$,
while $f(l)$ are smooth curves joining $u$ with $v$.
The proof of the theorem is now complete. \fbox{} \\

In order to prove Theorem~\ref{full-triangular}, we will use the
following general fact, of which we include a proof for completeness.

\begin{lemma} \label{general-fact}
If $\{ X_{\delta} \}, \{ Y_{\delta} \}$ (for $\delta > 0$), and $X$ are
random variables taking values in a complete,
separable metric space $S$ (whose $\sigma$-algebra is the Borel algebra)
with $\{ X_{\delta} \}$ and $\{ Y_{\delta} \}$ all defined on the same
probability space, then if $X_{\delta}$ converges in distribution to $X$ and
the metric distance between $X_{\delta}$ and $Y_{\delta}$ tends to zero
almost surely  as $\delta \to 0$, $Y_{\delta}$ also converges in distribution
to $X$.
\end{lemma}

\noindent {\bf Proof.} Since $X_{\delta}$ converges to $X$ in distribution,
the family $\{ X_{\delta} \}$ is relatively compact and therefore tight by
an application of Prohorov's Theorem (using the fact that $S$ is a
complete, separable metric space -- see, e.g., \cite{billingsley}).
Then, for any bounded, continuous, real function $f$ on $S$,
and for any $\varepsilon > 0$, there exists a compact set $\cal K$
such that $\int |f(X_{\delta})| I_{\{X_{\delta} \notin {\cal K}\}} dP < \varepsilon$
and $\int |f(Y_{\delta})| I_{\{X_{\delta} \notin {\cal K}\}} dP < \varepsilon$
for all $\delta$, where $I_{\{\cdot\}}$ is the indicator function and $P$ the
probability measure of the probability space of $\{ X_{\delta} \}$ and
$\{ Y_{\delta} \}$.
Thus, for small enough $\delta$,
\begin{equation}
|\int f(X_{\delta}) dP - \int f(Y_{\delta}) dP| <
\int |f(X_{\delta}) - f(Y_{\delta})| I_{\{X_{\delta} \in {\cal K}\}} dP + 2 \varepsilon
< 3 \varepsilon,
\end{equation}
where in the last inequality we use the absolute continuity of $f$ when
restricted to the compact set $\cal K$ and the fact that the metric distance
between $X_{\delta}$ and $Y_{\delta}$ goes to $0$ as $\delta \to 0$. \fbox{} \\

\noindent{\bf Proof of Theorem~\ref{full-triangular}.}
The theorem is an immediate consequence of Lemma~\ref{main-thm}
and Lemma~\ref{general-fact}.
It is enough to apply Lemma~\ref{general-fact} to the triplet
$\mu_{\delta}$, $\hat\mu_{\delta}$, $\mu$ (or, to be more precise,
to the random variables of which those are the distributions),
where $\mu$ is the full scaling limit of critical site percolation
on the triangular lattice~\cite{cn2,cn3}. \fbox{}

\bigskip

\noindent {\bf Acknowledgements.} 
%
%
Work on this paper began during a visit in 2002/2003 to
the Forschungs\-institut f\"ur Mathematik of ETH-Z\"urich,
and continued during a postdoc at EURANDOM, Eindhoven.
The author is grateful to the FIM and Alain-Sol Sznitman for their
kind hospitality, and acknowledges the support of a Marie Curie
Intra-European Fellowship (contract MEIF-CT-2003-500740) while
at EURANDOM.
The author also thanks Rob van den Berg and Frank den Hollander
for their advice, and Ronald Meester and the anonymous referees
for their comments on the presentation of the results.

\bigskip

\appendix
\refstepcounter{section}
\section*{Appendix \thesection: Matching Pairs of Lattices} \label{matching}

The proofs of the results of this paper concerning enhancement
percolation and the universality of critical exponents apply not
only to the square, triangular and hexagonal lattice, but to the
class of regular lattices considered in~\cite{kesten}. These are
infinite periodic graphs embedded in some suitable way in ${\mathbb
R}^2$. While we refer to Chapter~2 of~\cite{kesten} for the relevant
definitions not explicitly given here and all the details, we
explain below the general notion of matching pairs of graphs (or
lattices). Within this general framework, which includes in
particular the three lattices we are mainly interested in, we then
provide the proofs of the ``geometric" lemmas needed in
Section~\ref{proof-equivalence} in the the proof of
Lemma~\ref{equivalence}.

Let $\mathbb M$ be a regular, planar lattice embedded in ${\mathbb R}^2$
(a {\bf mosaic} in the language of Sykes and Essam~\cite{se}, and
Kesten~\cite{kesten}) such that each component $F$ of
${\mathbb R}^2 \setminus {\mathbb M}$ is bounded by a Jordan curve
(i.e., a simple, closed curve) made up of a finite number of edges
of $\mathbb M$.
$F$ is called a {\bf face} of $\mathbb M$, the edges delimiting
$F$ form its {\bf perimeter}, and the sites of $\mathbb M$ in the
perimeter of $F$ are its {\bf vertices}, which we denote by $V(F)$.
{\bf Close-packing} a face $F$ of $\mathbb M$ means adding an
edge between each pair of vertices of $F$ that do not already
share an edge.

Given a mosaic $\mathbb M$ and a subset $\mathfrak F$ of its collection
of faces, a lattice $\mathbb L$ is obtained from $\mathbb M$ by close-packing
all the faces in $\mathfrak F$, and ${\mathbb L}^*$ by close-packing all the
faces not in $\mathfrak F$. The pair ($\mathbb L$, ${\mathbb L}^*$) is a
{\bf matching pair of lattices}.

In the embedding of $\mathbb L$ and ${\mathbb L}^*$ one can choose
to draw the edges added to $\mathbb M$ when close-packing a face $F$
inside that same face. In a matching pair usually at least one of
the lattices $\mathbb L$ or ${\mathbb L}^*$ is not planar. Notice
that ${\mathfrak F} = \emptyset$ is allowed in the previous
definition; in that case $\mathbb L$ coincides with $\mathbb M$ and
${\mathbb L}^*$ is the close-packed version of $\mathbb M$.

The definitions of $\mathbb L$- and ${\mathbb L}^*$-adjacent,
$\mathbb L$- and $*$-path, $\mathbb L$- and $*$-connected, and
external (site) boundary are the same as in Section~\ref{lattices}.
We will also use the same notation for sites and paths as in that section.

It is important to observe (see, for example, \cite{se} and
Corollary~2.2~of~\cite{kesten}) that the external (site)
boundary of a nonempty, bounded, $\mathbb L$-connected set
$C$ of sites of $\mathbb L$ forms, together with the edges
between sites in the boundary, a self-avoiding $*$-loop $\lambda^*$
such that all the sites in $C$ belong to $\text{int}(\lambda^*)$.

Note that the square, triangular and hexagonal lattices considered
in the main part of the paper are of the type described above with
${\mathbb L} = {\mathbb M}$.

\begin{lemma} \label{faces}
Given a self-repelling $*$-loop $\lambda^*$ and an edge $\emph{\text{e}}=(x,y)$
in $\lambda^*$:
(i) if $\emph{\text{e}}$ belongs to the perimeters of faces $F_1$ and $F_2$ of
$\mathbb M$, then $F_1$ and $F_2$ belong one to $\emph{\text{int}}(\lambda^*)$
and the other to $\emph{\text{ext}}(\lambda^*)$;
(ii) if $\emph{\text{e}}$ is contained in the interior of face $F \notin {\mathfrak F}$,
then the two portions of $F$ separated by $\emph{\text{e}}$ belong one to
$\emph{\text{int}}(\lambda^*)$ and the other to $\emph{\text{ext}}(\lambda^*)$.
\end{lemma}

\noindent {\bf Proof.} (i) Since $\lambda^*$ is self-repelling,
sites in the perimeter of $F_1$ (resp., $F_2$) other than $x$ and $y$
can belong to $\lambda^*$ only if $F_1$ (resp., $F_2$) is not close-packed
in ${\mathbb L}^*$.
This implies that $\lambda^*$ does not cut through $F_1$, nor $F_2$.
Therefore the two faces, being on opposite sides of $\lambda^*$, belong
one to $\text{int}(\lambda^*)$ and the other to $\text{ext}(\lambda^*)$.

(ii) In this case, $F$ is of necessity close-packed in ${\mathbb L}^*$
and so no site in $V(F)$ other than $x$ and $y$ can be in $\lambda^*$.
Therefore the two portions of $F$, being on opposite sides of $\lambda^*$, belong
one to $\text{int}(\lambda^*)$ and the other to $\text{ext}(\lambda^*)$. \fbox{}

\begin{lemma} \label{neighbors}
If $\lambda^*$ is a self-repelling $*$-loop such that
$\emph{\text{int}}(\lambda^*) \cap V({\mathbb L})$ is not empty, then each
site in $\lambda^*$ has at least one
$\mathbb L$-neighbor in $\emph{\text{int}}(\lambda^*) \cap V({\mathbb L})$
and one in $\emph{\text{ext}}(\lambda^*) \cap V({\mathbb L})$.
\end{lemma}

\noindent {\bf Proof.} Consider two consecutive sites of
$\lambda^*$, $z_{j-1}$ and $z_j$.
We want to show that $z_j$ has at least one neighbor in each
of the two Jordan domains that make up ${\mathbb R}^2 \setminus \lambda^*$.
We first assume that the edge $(z_{j-1},z_j)$ belongs to
the perimeters of two faces of $\mathbb M$, $F_1$ and $F_2$.
By Lemma~\ref{faces}, one of the two faces, say $F_1$, must
be in $\text{int}(\lambda^*)$ and the other one, $F_2$, in
$\text{ext}(\lambda^*)$.

If $F_1$ belongs to $\mathfrak F$, it is close-packed
in $\mathbb L$ and therefore $z_j$ is $\mathbb L$-adjacent
to all the vertices of $F_1$.
On the other hand, not all the vertices of $F_1$ can belong to
$\lambda^*$, otherwise this would not be self-repelling.
This shows that $z_j$ has an $\mathbb L$-neighbor in $V(F_1)$
which is not in $\lambda^*$, thus it has an $\mathbb L$-neighbor
in $\text{int}(\lambda^*)$.
If $F_1$ does not belong to $\mathfrak F$, it is close-packed in
${\mathbb L}^*$ and therefore the sites of $V(F_1)$ other than
$z_{j-1}$ and $z_j$ do not belong to $\lambda^*$.
Then, $z_j$ has an $\mathbb L$-neighbor in $V(F_1)$ which does not
belong to $\lambda^*$ and thus belongs to $\text{int}(\lambda^*)$.
Arguments analogous to the ones above, but with $F_1$ replaced
by $F_2$, show that $z_j$ must have an $\mathbb L$-neighbor that
belongs to $\text{ext}(\lambda^*)$.

If the edge $(z_{j-1},z_j)$ is contained in the interior of a
face $F$, then of necessity $F$ is close-packed in ${\mathbb L}^*$
and no site of $V(F)$ other than $z_{j-1}$ and $z_j$ belongs
to $\lambda^*$.
By Lemma~\ref{faces}, $(z_{j-1},z_j)$ splits $F$ in two parts,
one contained in $\text{int}(\lambda^*)$ and the other in
$\text{ext}(\lambda^*)$.
The perimeter of each of those two parts contains $(z_{j-1},z_j)$
plus at least two more edges and one site $\mathbb L$-adjacent
to $z_j$.
Therefore $z_j$ has an $\mathbb L$-neighbor in $\text{int}(\lambda^*)$
and one in $\text{ext}(\lambda^*)$. \fbox{}

\begin{lemma} \label{partition}
A (finite) self-repelling $*$-loop $\lambda^*$ partitions
${\mathbb L} \setminus \lambda^*$ into two $\mathbb L$-connected
components, one bounded and the other unbounded.
\end{lemma}

\begin{remark} \label{remark-stronger}
Note that this is a stronger statement than saying that $\lambda^*$,
being a Jordan curve, partitions ${\mathbb R}^2 \setminus \lambda^*$
in two components.
In fact, we are claiming that the subsets of $\mathbb L$ contained
in $\emph{\text{int}}(\lambda^*)$ and $\emph{\text{ext}}(\lambda^*)$
are $\mathbb L$-connected, and this can be false if $\lambda^*$ is
not self-repelling.
\end{remark}

\noindent {\bf Proof.} As explained in Remark~\ref{remark-stronger},
we need to prove that $\text{int}(\lambda^*) \cap {\mathbb L}$ and
$\text{ext}(\lambda^*) \cap {\mathbb L}$ are $\mathbb L$-connected.
We only give the proof of this fact for $\text{int}(\lambda^*) \cap {\mathbb L}$
(that for $\text{ext}(\lambda^*) \cap {\mathbb L}$ being analogous) and split
it in two parts.
We will first show that, for each site $z_j \in \lambda^*$, any two sites
belonging to the set ${\cal N}_{\mathbb L}^{\lambda^*}(z_j)$ of $\mathbb L$-neighbors
of $z_j$ contained in $\text{int}(\lambda^*)$ can be joined by an $\mathbb L$-path
completely contained in $\text{int}(\lambda^*)$.
Then, for any two successive sites $z_{j-1}$ and $z_j$ in $\lambda^*$, we will
show that they have two neighbors, $z'_{j-1}$ and $z'_j$ respectively, which
belong to $\text{int}(\lambda^*) \cap V({\mathbb L})$ and such that there is an
$\mathbb L$-path joining them and completely contained in $\text{int}(\lambda^*)$.

These two facts prove that any site in ${\cal N}_{\mathbb L}^{\lambda^*}(z_{j-1})$
can be joined to any site in ${\cal N}_{\mathbb L}^{\lambda^*}(z_j)$ by an
$\mathbb L$-path completely contained in $\text{int}(\lambda^*)$.
Therefore, for each pair of sites $z_i$ and $z_j$ in $\lambda^*$,
any site in ${\cal N}_{\mathbb L}^{\lambda^*}(z_i)$ can be joined
to any site in ${\cal N}_{\mathbb L}^{\lambda^*}(z_j)$ by an $\mathbb L$-path
completely contained in $\text{int}(\lambda^*)$.
This clearly proves the claim, since each $x \in \text{int}(\lambda^*) \cap V({\mathbb L})$
is either an $\mathbb L$-neighbor of a site in $\lambda^*$ or can be joined to one
by an $\mathbb L$-path completely contained in $\text{int}(\lambda^*)$.

We will now prove the first claim about ${\cal N}_{\mathbb L}^{\lambda^*}(z_j)$,
for $z_j \in \lambda^*$.
By Lemma~\ref{neighbors}, $z_j$ has at least one $\mathbb L$-neighbor
in $\text{int}(\lambda^*) \cap {\mathbb L}$.
If it has exactly one such $\mathbb L$-neighbor, there is nothing to
prove, so we assume that $z_j$ has at least two $\mathbb L$-neighbors
in $\text{int}(\lambda^*) \cap {\mathbb L}$.
We first show that there is an $\mathbb L$-path, contained
in $\text{int}(\lambda^*)$, joining any two elements of
${\cal N}_{\mathbb L}^{\lambda^*}(z_j)$ whenever they are
vertices of the same face $F$.
Consider two such sites $x$ and $y$ in ${\cal N}_{\mathbb L}^{\lambda^*}(z_j)$.
If $F$ is close-packed in $\mathbb L$, $x$ and $y$ are automatically
adjacent in $\mathbb L$.
If $F$ is not close-packed in $\mathbb L$, $x$ and $y$ are the only
$\mathbb L$-neighbors of $z_j$ belonging to $V(F)$.
Let $z_{j-1}$ and $z_{j+1}$ be the sites that come before and after
$z_j$ in $\lambda^*$.
They cannot both belong to $V(F)$, otherwise $\lambda^*$ would not be
self-repelling.
If only one, say $z_{j-1}$ belongs to $V(F)$, then the edge $(z_{j-1},z_j)$
that cuts through $F$ belongs to $\lambda^*$ and, by Lemma~\ref{faces} (ii),
it cannot be the case that $x$ and $y$ are both in $\text{int}(\lambda^*)$,
contradicting our hypothesis.
Thus, neither $z_{j-1}$ nor $z_{j+1}$ belongs to $V(F)$, which implies
that, since $F$ is close-packed in ${\mathbb L}^*$, no site of
$V(F)$ other than $z_j$ can belong to $\lambda^*$ (or else this would
not be self-repelling).
Therefore, all the sites in $V(F) \setminus {\{ z_j \}}$ are inside
$\text{int}(\lambda^*)$ and can be used to join $x$ with $y$ with an
$\mathbb L$-path contained in $\text{int}(\lambda^*)$.

If $x$ and $y$ are both in ${\cal N}_{\mathbb L}^{\lambda^*}(z_j)$
but they are not vertices of the same face, then we are going to
prove that there exists a sequence $z^0_j, z^1_j, \ldots, z^k_j$,
with $z^0_j=x$ and $z^k_j=y$, of elements of
${\cal N}_{\mathbb L}^{\lambda^*}(z_j)$ such that for each
$i \in 1, \ldots, k$ there is a face $F_i$ of $\mathbb M$ with
$z^{i-1}_j, z^i_j \in V(F_i)$, so that we can use again the result
just above to find an $\mathbb L$-path inside $\text{int}(\lambda^*)$
joining $x$ with $y$.

To see this, consider all the edges of ${\mathbb L}^*$ incident on $z_j$.
Since $\lambda^*$ is self-repelling, only two of those edges and the
two $*$-neighbors of $z_j$ that they are incident on can belong to $\lambda^*$.
Then, the remaining edges and $*$-neighbors of $z_j$ are divided in two groups,
those that lie to the right of $\lambda^*$ and those that lie to its left.
The edges and sites in one group are contained in $\text{int}(\lambda^*)$,
those in the other group are contained in $\text{ext}(\lambda^*)$.
$x$ and $y$ belong to the group in $\text{int}(\lambda^*)$.
Since we are assuming that $x$ and $y$ are not vertices of the same face,
spanning the wedge from the edge $(x, z_j)$ to the edge $(y, z_j)$, or vice
versa, in such a way as to remain inside $\text{int}(\lambda^*)$, one must
encounter one or more edges of $\mathbb M$ incident on $z_j$.
Assuming (without loss of generality) that the right way to span the wedge
is from $(x, z_j)$ to $(y, z_j)$, we order those edges according to the
order in which they are encountered moving from $(x, z_j)$ to $(y, z_j)$,
including $(x, z_j)$ and $(y, z_j)$, and call them, respectively,
$(z^0_j, z_j), (z^1_j, z_j), \ldots, (z^k_j, z_j)$, with $z^0_j=x$ and
$z^k_j=y$.
The sequence $z^0_j, z^1_j, \ldots, z^k_j$ is what we are after, and
with this the claim that any two sites in ${\cal N}_{\mathbb L}^{\lambda^*}(z_j)$
can be connected by an $\mathbb L$-path completely contained in
$\text{int}(\lambda^*)$ is proved.

In order to conclude the proof, we will now show that, given $z_{j-1}$ and $z_j$
in $\lambda^*$, one can pick a site $z'_{j-1}$ from ${\cal N}_{\mathbb L}^{\lambda^*}(z_{j-1})$
and a site $z'_j$ from ${\cal N}_{\mathbb L}^{\lambda^*}(z_j)$ such that there is an
$\mathbb L$-path joining $z'_{j-1}$ and $z'_j$, and completely contained in $\text{int}(\lambda^*)$.

We first assume that the edge $(z_{j-1},z_j)$ belongs to the perimeters
of two faces, $F_1$ and $F_2$, of $\mathbb M$.
By Lemma~\ref{faces}, one of these two faces, say $F_1$, is contained
inside $\text{int}(\lambda^*)$.
If $F_1$ belongs to $\mathfrak F$, it is close-packed
in $\mathbb L$ and therefore $z_{j-1}$ and $z_j$ are
$\mathbb L$-adjacent to all the other vertices of $F_1$.
On the other hand, not all the vertices of $F_1$ can belong
to $\lambda^*$, otherwise this would not be self-repelling.
Thus, $z_{j-1}$ and $z_j$ have a common $\mathbb L$-neighbor
in $V(F_1)$ which does not belong to $\lambda^*$ and is therefore
contained in $\text{int}(\lambda^*)$.
This common neighbor is our choice for both $z'_{j-1}$ and $z'_j$.

If $F_1$ does not belong to $\mathfrak F$, it is close-packed in
${\mathbb L}^*$ and therefore the sites of $V(F_1)$ other than
$z_{j-1}$ and $z_j$ do not belong to $\lambda^*$.
Then, $z_{j-1}$ and $z_j$ have $\mathbb L$-neighbors $z'_{j-1}$ and $z'_j$
in $V(F_1)$ which do not belong to $\lambda^*$ and are contained in
$\text{int}(\lambda^*)$.
$z'_{j-1}$ and $z'_j$ may coincide or be $\mathbb L$-adjacent; if not,
they nonetheless belong to the perimeter of the same face $F_1$ and
can therefore be joined by an $\mathbb L$-path contained in
$\text{int}(\lambda^*)$ that uses the other sites of $V(F_1)$.

If the edge $(z_{j-1},z_j)$ is contained in the interior of a
face $F$, then of necessity $F$ is close-packed in ${\mathbb L}^*$
and no site in $V(F)$ other than $z_{j-1}$ and $z_j$ belongs
to $\lambda^*$ .
By Lemma~\ref{faces}, $(z_{j-1},z_j)$ splits $F$ in two parts,
one contained in $\text{int}(\lambda^*)$ and the other in
$\text{ext}(\lambda^*)$.
The perimeter of the portion in $\text{int}(\lambda^*)$ contains
$(z_{j-1},z_j)$ plus at least two more edges and two sites,
$z'_{j-1}$ and $z'_j$ (which may coincide), $\mathbb L$-adjacent
to $z_{j-1}$ and $z_j$ respectively.
$z'_{j-1}$ and $z'_j$ either coincide, or are $\mathbb L$-adjacent,
or can be joined by an $\mathbb L$-path that uses vertices of $F$
in $\text{int}(\lambda^*)$. \fbox{} \\

\end{document}